\documentstyle[12pt,psfig]{article}

\newcommand{\resection}[1]{\setcounter{equation}{0}\section{#1}}
\newcommand{\appsection}{\addtocounter{section}{1} \setcounter{equation}{0}
                         \section*{Appendix \Alph{section}}}

\newcommand{\EQ}{\begin{equation}}
\newcommand{\EN}{\end{equation}}
\newcommand{\bea}{\begin{eqnarray}}
\newcommand{\eea}{\end{eqnarray}}
\newcommand{\hs}{\hspace{0.1cm}}

\newcommand{\st}{\stackrel}
\newcommand{\th}{\theta}
\newcommand{\al}{\alpha}

\newcommand{\goto}{\rightarrow}
\newcommand{\lab}{\label}

\begin{document}
\setcounter{page}{0}
\topmargin 0pt
\oddsidemargin 5mm
\renewcommand{\thefootnote}{\arabic{footnote}}
\newpage
\setcounter{page}{0}
\begin{titlepage}
\begin{flushright}
OUTP-95-385\\
ISAS/EP/95/78
\end{flushright}
\vspace{0.5cm}
\begin{center}
{\large {\bf The Spin-Spin Correlation Function in the Two-Dimensional \\
Ising Model in a Magnetic Field at $T=T_{c}$}}\\
\vspace{1.8cm}
{\large G. Delfino} \\
\vspace{0.5cm}
{\em Theoretical Physics, University of Oxford}\\
{\em 1 Keble Road, Oxford OX1 3NP, United Kingdom}\\
\vspace{0.5cm}
{\large and}\\
\vspace{0.5cm}
{\large G. Mussardo}\\
\vspace{0.5cm}
{\em International School for Advanced Studies,\\
and \\
Istituto Nazionale di Fisica Nucleare\\
34014 Trieste, Italy}\\

\end{center}
\vspace{1.2cm}

\renewcommand{\thefootnote}{\arabic{footnote}}
\setcounter{footnote}{0}

\begin{abstract}
\noindent
The form factor bootstrap approach is used to compute the exact contributions
in the large distance expansion of the correlation function
$<\sigma(x) \sigma(0)>$ of the two-dimensional Ising model in a magnetic
field at $T=T_{c}$. The matrix elements of the magnetization operator
$\sigma(x)$ present a rich analytic structure induced by the (multi)
scattering processes of the eight massive particles of the model. The spectral
representation series  has a fast rate of convergence and perfectly agrees
with the numerical determination of the correlation function.
\end{abstract}

\vspace{.3cm}

\end{titlepage}

\newpage
\noindent
\resection{Introduction}

Over the past few years, considerable progress has been made in the use
of conformal invariance methods and scattering theory for the understanding of
the critical points and the nearby scaling region of two-dimensional
statistical
models (see, for instance \cite{ISZ,GM}). At the critical points,
the correlation functions of the statistical models fall into a
scale-invariant regime and their computation may be achieved by solving the
linear differential equations obtained by the representation theory of the
infinite-dimensional conformal symmetry \cite{BPZ}. The situation is different
away from criticality. The scaling region may be described in terms
of the relevant deformations of the fixed point actions. These
deformations destroy the long-range fluctuations of the critical point
and the associated quantum field theories are usually massive. If an infinite
number of conserved charges survive the deformation of the critical point
action, the corresponding QFT can be efficiently characterized (on-shell) by
the relativistic scattering processes of the massive excitations. In this case,
the integrals of motion severely restrict the bound state structure
of the theory and force the $S$-matrix to be elastic and factorizable
\cite{Zam,ZZ}. Once the exact $S$-matrix of a model is known, one can proceed
further and investigate the off-shell behaviour of the theory by means of the
Form Factor approach \cite{KW,Smirnov}. This consists in computing matrix
elements of the local fields on the set of asymptotic states and
reconstructing their correlation functions in terms of the spectral
density representation. As argued in \cite{CMpol}, and also confirmed by
the explicit solution of several two-dimensional massive
QFT \cite{YZ,JLG,ZamYL,DM,FMS}, a general
property of the spectral series of massive integrable QFT is their very fast
rate of convergence for {\em all} distance scales. This important quality
of the spectral representation series may be regarded as the key point to
the success of the Form Factor approach.

The aim of this paper is to apply the Form Factor approach and compute the
spin-spin correlation function $G(x) = <\sigma(x) \sigma(0) >$
of the Ising model in a magnetic field $h$ at $T =T_c$ (in the sequel,
this model will be referred to as IMMF). For the other integrable deformation
of the critical Ising model, i.e. thermal, the spin-spin correlation function
has been exactly determined in \cite{MW,McBook} and this result,
together with its remarkable connection with the Painleve' functions
\cite{Painleve,BB}, may be regarded as one of the main accomplishments
in statistical mechanics. On the other hand, very little is known about the
spin-spin correlation function for $h\neq 0$ at $T=T_c$ whose determination
has been a long-standing problem of statistical mechanics. As we will show in
this paper, the computation of $G(x)$ can now be approached analytically,
thanks to the scattering formulation of the model proposed by Zamolodchikov
\cite{Zam}. Apart from the importance of computing $G(x)$ itself, there are
other related theoretical issues which render this calculation
instructive.

The first issue concerns the rich structure of bound states and higher-order
poles of the $S$-matrix. Higher-order poles of the $S$-matrix of
two-dimensional field theories can be naturally interpreted as singularities
associated to multi-scattering processes \cite{ColT,BCDS,ChM}. On this basis,
one would expect a similar hierarchy of higher-order poles in the form
factors as well.
There is, however, an important difference between the pole structures of the
$S$-matrix and those of the form factors. In fact, the $S$-matrix contains
{\em simultaneously} information about the $s$-channel as well as the
$u$-channel. Correspondingly, the poles of the $S$-matrix are always
arranged in pairs, with their positions located in crossing symmetrical
points. This is in concordance with the two different ways of looking
at a scattering process, i.e. in the direct or in the crossed channel. On the
contrary, the $u$-variable plays no role in the calculation of the
form factors as these only depend on the $s$-variables of all
subchannels of the asymptotic states on which the matrix elements are
computed. The absence of $u$-channel singularities in the form factors
implies that their analytic structure may be different from that of the
$S$-matrix and it is therefore an interesting problem to understand how
the higher-order poles enter the form factor calculation.

The second theoretical issue that emerges from the computation of the
spin-spin correlation function in the IMMF is a careful reconsideration
of the so-called {\em minimality prescription}, which is usually invoked
for computing the form factors. Briefly stated, this consists in
assuming that the form factors have the minimal analytic structure,
compatible with the nature of the operator and the bound state pattern
of the scattering theory. So far, this prescription has been
successfully applied to solve several two-dimensional QFT as, for instance,
those of refs.\,\cite{YZ,JLG,ZamYL,DM,FMS}, despite the fact
that the theoretical reason of its validity was lacking. One of the basic
motivations why minimality is often adopted is because frequently
the asymptotic behaviour of the matrix elements is not easy to determine.
In this paper, we will present a simple argument which will allow us
to place a quite restrictive upper bound on the high energy
limit of form factors. Using this criterion, one can see that the minimality
prescription is violated in the IMMF and therefore extra polynomials in the
variable $s$ have to be included in the matrix elements of the field
$\sigma(x)$.
These polynomials, nevertheless, can be uniquely determined since
the scattering theory always provides a sufficient number
of constraints. As a matter of fact, the equations which fix
the extra polynomials are usually overdetermined and this gives rise
to self-consistency conditions which are indeed fulfilled
by the IMMF form factors.

The last point we would like to mention is the successful comparison of
our theoretical determination of $<\sigma(x) \sigma(0)>$ with numerical
simulations. These simulations have been carried out in the last few years by
two different groups \cite{R,LR,D} and in particular, a collection of
high-precision numerical estimates of the spin-spin correlation function of
the IMMF, for different values of the magnetic field and different sizes of
the lattice, can be found in \cite{LR}. Although there is no doubt about the
actual existence of higher-mass particles, which can be easily extracted by
transfer matrix diagonalization of the IMMF \cite{numspectrum} or directly
from the lattice \cite{nienhuis}, it was quite difficult to see the presence
of these higher-mass states from the analysis of the numerical
data of the spin-spin correlation function. Namely, a best fit of the two-point
function seemed to be compatible with the only contribution of the fundamental
particle \cite{R}, a result which appears intriguing.
In fact, from a theoretical point of view there is no reason
why decoupling of higher-mass particles should occur in the spin-spin
correlator
since the IMMF has apparently no selection rules related to any symmetry.
Indeed, as we will see, the magnetization field $\sigma(x)$ couples to
{\em all} particles of the theory and it is only the small values of the
relative couplings which could be responsible for a possible misleading
interpretation of the numerical results.

The paper is organized as follows. In Section 2, we briefly outline
the main features of scattering theory of the IMMF, the exact mass
spectrum of the model and some other quantities which will prove
useful in the sequel of the paper. In Section 3, we analyse the Form
Factor approach and the properties of the spectral representation
series of the correlation functions. In Section 4, a general
constraint on the asymptotic behaviour of the form factors is introduced
and applied to the computation of the matrix elements of the magnetization
operator $\sigma(x)$ of the IMMF. We also discuss the
occurrence and the interpretation of higher order poles in the form factors.
Comparison of our theoretical results versus numerical simulations,
as well as the saturation of the sum-rules satisfied by the
spin-spin correlation function, are presented in Section 5.
The paper also contains three appendices. In Appendix A we
discuss the general bootstrap approach to the computation of the form
factors of the model. Appendix B gathers useful mathematical formulas
to deal with the monodromy properties of the matrix elements.
Appendix C presents specific examples of matrix elements with
singularities associated to higher order poles in the scattering amplitudes.

\resection{Scattering theory}

In the continuum limit, the IMMF may be regarded as a perturbed CFT.
At the critical point, ($T=T_c$ and $h=0$), the Ising model is
described by the conformal minimal model ${\cal M}_{3,4}$ with central charge
$C = \frac{1}{2}$ \cite{BPZ}. There are
three conformal families in the model, those of the identity, magnetization
and energy operators, denoted respectively as $[I = (0,0)]$,
$\left[\sigma = \left(\frac{1}{16},\frac{1}{16}\right)\right]$ and
$\left[\epsilon = \left(\frac{1}{2},\frac{1}{2}\right)\right]$, where the
numbers $(\Delta_i,\Delta_i)$ in the round brackets are their conformal
weights. We can move the system away from criticality by modifying the action
${\cal A}_0$ of the critical point as
\EQ
{\cal A} = {\cal A}_0 + h \int d^2x \, \sigma(x) \,\,\,.
\label{action}
\EN
For small values of the magnetic field, the system is still at $T = T_c$.
However, the coupling to the magnetic field induces a mass scale $M(h)$ in the
problem and destroys the long-range fluctuations of the critical point.
Using the action (\ref{action}), or the corresponding lattice Hamiltonian,
one can in principle analyse the properties of the model in terms of a
perturbative expansion in $h$ \cite{McCoy,Dotsenko}. There are, however,
some important questions which cannot be easily addressed by using such
perturbative formulation, as, for instance, the determination of the exact
mass spectrum of the model or the computation of its correlation functions.
In order to answer such questions, we have to rely on a non-perturbative
approach that exploits the most important feature of the dynamics of
the model, namely its integrability. The IMMF has, in fact, an infinite
number of conserved charges of spin $s=1,7,11,13,17,19,23,29$ (mod $30$).
As shown in a remarkable paper by Zamolodchikov \cite{Zam}, the existence
of these conserved charges allows one to define a self-consistent
scattering theory which provides an alternative route to the perturbative
formulation. Let us briefly outline the main results discussed
in \cite{Zam}.

The scattering theory which describes the scaling limit of the IMMF
contains eight different species of self-conjugated particles
$A_{a}$, $a=1,\ldots,8$ with masses
\bea
m_1 &=& M(h)\,, \nonumber \\
m_2 &=& 2 m_1 \cos\frac{\pi}{5} = (1.6180339887..) \,m_1\,,\nonumber\\
m_3 &=& 2 m_1 \cos\frac{\pi}{30} = (1.9890437907..) \,m_1\,,\nonumber\\
m_4 &=& 2 m_2 \cos\frac{7\pi}{30} = (2.4048671724..) \,m_1\,,\nonumber \\
m_5 &=& 2 m_2 \cos\frac{2\pi}{15} = (2.9562952015..) \,m_1\,,\\
m_6 &=& 2 m_2 \cos\frac{\pi}{30} = (3.2183404585..) \,m_1\,,\nonumber\\
m_7 &=& 4 m_2 \cos\frac{\pi}{5}\cos\frac{7\pi}{30} = (3.8911568233..) \,m_1\,,
\nonumber\\
m_8 &=& 4 m_2 \cos\frac{\pi}{5}\cos\frac{2\pi}{15} = (4.7833861168..) \,m_1\,,
\nonumber
\eea
Within the standard CFT normalization of the magnetization operator,
fixed by the equation
\EQ
<\sigma(x)\sigma(0)>=\frac{1}{\,\,|x|^{\frac{1}{4}}}\,\,,\hspace{1cm}|x|\goto
0\lab{uv}
\EN
the overall mass scale $M(h)$ has been exactly determined in \cite{Fateev},
\EQ
M(h) \,=\, {\cal C} \, h^{\frac{8}{15}}
\label{fat}
\EN
where
\EQ
{\cal C} \,=\,
 \frac{4 \sin\frac{\pi}{5} \Gamma\left(\frac{1}{5}\right)}
{\Gamma\left(\frac{2}{3}\right) \Gamma\left(\frac{8}{15}\right)}
\left(\frac{4 \pi^2 \Gamma\left(\frac{3}{4}\right)
\Gamma^2\left(\frac{13}{16}\right)}{\Gamma\left(\frac{1}{4}\right)
\Gamma^2\left(\frac{3}{16}\right)}\right)^{\frac{4}{5}} \,
 \,=\, 4.40490858...
\EN
The scattering processes in which the eight particles are involved
are completely elastic (the final state contains exactly the same
particles as the initial one with unchanged momenta) and, due to the
factorization of multiparticle scattering amplitudes induced by integrability,
they are entirely characterised by the two-particle amplitudes $S_{ab}$
(Figure 1). These are functions of the relativistic invariant Mandelstam
variable $s=(p_a + p_b)^2$ or, equivalently, of $u=(p_a - p_b)^2$.
$S_{ab}$ has a branch root singularity in the variable $s$ at the threshold
$s=(m_a + m_b)^2$. By crossing symmetry, an analogous branch point also
appears at the threshold of the $u$-channel, namely at
$s=(m_a - m_b)^2$ (Figure 2). Those are the only branch cuts of the
$S$-matrix, due to the elastic nature of the scattering processes.
The other possible singularities of the scattering amplitudes $S_{ab}$
are simple and higher-order poles in the interval
$ (m_a - m_b)^2 < s < (m_a + m_b)^2$ which are related to the bound state
structure.

An important simplification in the analysis of the analytic structure of the
$S$-matrix comes from the parameterization of the external momenta
in terms of the rapidity variable $\th$, i.e.
$p_{a}^{0}=m_{a}\cosh\th_{a}$, $p_{a}^{1}=m_{a}\sinh\th_{a}$.
The mapping $ s(\th_{ab}) = m_a^2 + m_b^2 + 2 m_a m_b \cosh\th_{ab}$,
where $\th_{ab}=\th_a-\th_b$, or equivalently $ u(\th_{ab}) =
s(i \pi -\th_{ab}) = m_a^2 + m_b^2 - 2 m_a m_b \cosh\th_{ab}$,
transforms the amplitudes $S_{ab}$ into meromorphic functions
$S_{ab}(\th_{ab})$, which satisfy the equations
\EQ
S_{ab}(\th)S_{ab}(-\th)=1\,\,\,;
\label{unitarity}
\EN
\EQ
S_{ab}(\th) = S_{ab}(i\pi-\th)\,\, ,
\label{crossing}
\EN
expressing the unitarity and the crossing symmetry of the theory,
respectively. The simple poles of $S_{ab}(\th)$ with positive residues
are related to bound state propagation in the $s$-channel, as shown in
Figure 3, whereas those with negative residues are associated to bound
states in the $u$-channel. Suppose that the particle $A_c$ with
mass squared
\EQ
m_c^2 = m_a^2 + m_b^2 + 2m_a m_b \cos u_{ab}^c,
\hspace{1cm} u_{ab}^c\in(0,\pi)
\label{triangle}
\EN
is a stable bound state in the $s$-channel of the particles
$A_{a}$ and $A_{b}$. In the vicinity of the resonance angle
$\th = i u_{ab}^c$, the amplitude becomes
\EQ
S_{ab}(\th\simeq
iu_{ab}^c)\simeq\frac{i\left(\Gamma_{ab}^c\right)^2}{\th-iu_{ab}^c}\,\,,
\lab{spole}
\EN
with $\Gamma_{ab}^c$ denoting the three-particle coupling\footnote{Note that
the $S$-matrix cannot determine the three-particle couplings but only their
square. This results in an ambiguity for the sign of all $\Gamma_{ab}^c$ which
can be solved by defining a consistent set of form factors.}. Since the
bootstrap principle gives the possibility to consider the bound states
on the same footing as the asymptotic states, the amplitudes $S_{ab}$ are
related each other by the functional equations \cite{Zam,ZZ}
\EQ
S_{il}(\th) \,=\,S_{ij}(\th + i \overline u_{jl}^k)\,
S_{ik}(\th - i \overline u_{lk}^j) \,\, ,
\label{bootstrap}
\EN
($\overline u_{ab}^c \equiv \pi - u_{ab}^c$). For the IMMF, the solution of
eqs. (\ref{unitarity}), (\ref{crossing}) and (\ref{bootstrap}) are given by
\EQ
S_{ab}(\th) = \prod_{\alpha\in{\cal A}_{ab}}
\left(f_{\alpha}(\th)\right)^{p_\alpha}
\label{smatr}
\EN
where
\EQ
f_{\al}(\th) \equiv \frac{\tanh\frac{1}{2}\left(\th+i\pi
\al\right)}
     {\tanh\frac{1}{2}\left(\th-i\pi \al\right)}\,\,\,.
\label{elesmatrix}
\EN
The sets of numbers ${\cal A}_{ab}$ and their multiplicities
$p_\alpha$, specifying the amplitudes (\ref{smatr}), can be
found in Table 1\footnote{Note that the numbers ${\cal A}_{ab}$
of Table 1 should be read in units of $\frac{1}{30}$.}. In order to
correctly interpret this collection of data, notice that the functions
$f_{\alpha}(\th)$ satisfy the equation $f_{\alpha}(\th)=f_{1 -\alpha}(\th)$
as well as $f_{\alpha}(\th) = f_{\alpha}(i \pi -\th)$. Hence, they have
two poles located at the crossing symmetrical positions $\th = i \pi \alpha$
and $\th = i \pi (1-\alpha)$. Therefore, the poles of the
$S$-matrix of the IMMF will always appear in pairs.

Applying eq. (\ref{spole}) to the simple poles with positive residues
at $\th = \frac{2 \pi i}{3}$, $\th = \frac{2 \pi i}{5}$ and
$\th = \frac{i\pi}{15}$ in $S_{11}(\th)$ (related to the bound
states $A_1$, $A_2$ and $A_3$ respectively), we can extract
\[
\Gamma_{11}^1 \,=\,\sqrt{\frac{2 \tan \frac{2\pi}{3} \,\tan\frac{8\pi}{15}
\,\tan\frac{11 \pi}{30}}
{\tan\frac{2\pi}{15} \,\tan\frac{3\pi}{10}}} \,=\,10.990883..
\]
\EQ
\Gamma_{11}^2 \,=\, \sqrt{\frac{2 \tan \frac{2\pi}{5} \,\tan\frac{8\pi}{15}
\,\tan\frac{7 \pi}{30}}
{\tan\frac{2\pi}{15} \,\tan\frac{\pi}{6}}} \,=\,14.322681..
\EN
\[
\Gamma_{11}^3 \,=\,\sqrt{\frac{2 \tan \frac{\pi}{15} \,\tan\frac{11\pi}{30}
\,\tan\frac{7 \pi}{30}}
{\tan\frac{\pi}{6} \,\tan\frac{3\pi}{10}}} \,=\,1.0401363..
\]
Other three-particle couplings can be obtained similarly.
In addition to simple poles, the $S$-matrix of the IMMF presents higher-order
poles due to multi-scattering processes\footnote{A detailed analysis of the
nature of higher-order poles in the $S$-matrix of the ($1 + 1$)
integrable theories can be found in \cite{ColT,BCDS,ChM}.}.
The odd order poles correspond to bound state poles while those of even order
do not. Their appearance is an unavoidable consequence of the iterative
application of the bootstrap equations (\ref{bootstrap}). In relation with
the calculation of the Form Factors, some of these poles
will be considered in Section 4 and in Appendix C.

\resection{Correlation Functions and Form Factors}

Once the spectrum and the $S$-matrix are known, we can investigate the
off-shell behaviour of the theory. We can compute the two-point (as well as
higher-point) correlation functions of the model through the unitarity sum
(Figure 4)
\begin{eqnarray}
& & <\Phi(x) \Phi(0)> = \sum_{n=0}^{\infty} \int_{\th_1 >\th_2 \ldots
>\th_n} \frac{d\th_1}{2\pi} \cdots \frac{d\th_n}{2\pi}
\label{spectral} \\
& & \,\,\,\,\,\,|<0|\Phi(0)|A_{a_1}(\th_1) \cdots A_{a_n}(\th_n)>|^2
e^{-|x| \sum_{k=1}^n m_k \cosh\th_k} \nonumber
\end{eqnarray}
Basic quantities of this approach are the Form Factors (FF), i.e. the matrix
elements of the local operators $\Phi$ on the asymptotic states
(Figure 5), defined as
\EQ
F^{\Phi}_{a_1,\ldots ,a_n}(\th_1,\ldots,\th_n) = <0|
\Phi(0)|A_{a_1}(\th_1),\ldots,A_{a_n}(\th_n)>\,\,\,.
\label{form}
\EN
A detailed discussion of the form factor approach and the mathematical
properties of the matrix elements (\ref{form}) can be found in
\cite{KW,Smirnov}. Here we simply consider the basic equations we need for
our discussion.

For a scalar operator $\Phi(x)$, relativistic invariance requires that its
form factors depend only on the rapidity differences $\th_i-\th_j$. The
elasticity of the scattering processes, together with the crossing symmetry
and the completeness relation of the asymptotical states, permit
to derive the following monodromy equations satisfied by the FF
\EQ
\begin{array}{lll}
F^{\Phi}_{a_1,..,a_i,a_{i+1},.. a_n}(\th_1,..,\th_i, \th_{i+1}, ..
, \th_{n}) &=& \,S_{a_i a_{i+1}}(\th_i-\th_{i+1}) \,
F^{\Phi}_{a_1,..a_{i+1},a_i,..a_n}(\th_1,..,\th_{i+1},\th_i ,.., \th_{n})
\,\, ,\\
F^{\Phi}_{a_1,a_2,\ldots a_n}(\th_1+2 \pi i, \dots, \th_{n-1}, \th_{n} ) &=&
F^{\Phi}_{a_2,a_3,\ldots,a_n,a_1}(\th_2 ,\ldots,\th_{n}, \th_1)
\,\, .
\end{array}
\label{permu1}\\
\EN
In terms of the $s$-variable of the channel
$(A_{a_i}\, A_{a_{j}})$, the first equation in (\ref{permu1})
implies that in the form factors there is a branch cut in the $s$-plane
extending from $s = (m_{a_i} + m_{a_j})^2$ to $s = \infty$.
This is similar to what happens in the $S$-matrix. However,
a difference between the analytic structure of the FF and the $S$-matrix
comes from the second equation, which on the contrary shows that the FF
do not have a $u$-channel cut extending from $s=-\infty$
to $s=(m_{a_i} -m_{a_j})^2$.

Apart from these monodromy properties, the FF are expected to have poles
induced by the singularities of the $S$-matrix. A particular role is
played by the simple poles. Among them, we can select two different
classes which admit a natural particle interpretation. The first type of
simple poles are the so-called kinematical poles related to the
annihilation processes of a pair of particle and anti-particle states.
These singularities are located at $\th_{a}-\th_{\overline a} = i\pi$
and for the corresponding residue we have
\EQ
-i\lim_{\tilde\th \rightarrow \th}
(\tilde\th - \th)
F^{\Phi}_{a,\overline a,a_1,\ldots,a_n}
(\tilde\th + i\pi,\th,\th_1,\ldots,\th_n) =
\left(1 - \prod_{j=1}^{n} S_{a,a_j}(\th -\th_j)\right)\,
F^{\Phi}_{a_1,\ldots,a_n}(\th_1,\ldots,\th_{n})  . \label{recursive}
\EN
This equation can be graphically interpreted as an interference process
due to the two different kinematical pictures drawn in Fig.\,6.

A second class of simple poles is related to the presence of bound states
appearing as simple poles in the $S$-matrix. If $\th = i u_{ab}^c$ and
$\Gamma_{ab}^c$ are the resonance angle and the three-particle
coupling of the fusion $A_a \times A_b \rightarrow A_c$ respectively, then
FF involving the particles $A_a$ and $A_b$ will also have a pole at
$\th = i u_{ab}^c$, with the residue given by (Fig.\,7)
\EQ
-i \lim_{\th_{ab} \rightarrow i u_{ab}^c}
(\th_{ab} - i u_{ab}^c) \,
F^{\Phi}_{a,b,a_i,\ldots,a_n}
\left(\th_a, \th_b ,\th_1,\ldots,\th_n\right) =
\Gamma_{ab}^c \,
F^{\Phi}_{c,a_i,\ldots,a_n}\left(\th_c,\th_1,\ldots,\th_n\right)
\label{recurb}
\EN
where
$\th_c = (\th_a \overline u_{bc}^a + \th_b \overline u_{ac}^b)/u_{ab}^c$.

In general, the FF may also present simple poles which do not fall into the
two classes above. In addition, they may also have higher-order poles and
indeed, their analytic structure may be quite complicated. We will
came back to this point in Section 4 where the specific example of the
IMMF will be discussed.

Although eqs.\,(\ref{recursive}) and (\ref{recurb}) do not exhaust all pole
information, nevertheless they induce a recursive structure in the
space of the FF, which may be useful for their determination\footnote{
An important general aspect of the Form Factor approach which is worth
mentioning is that the validity of eqs. (\ref{permu1}), (\ref{recursive})
and (\ref{recurb}) do not rely on the choice of any specific local operator
$\Phi(x)$. This observation, originally presented in \cite{JLG}, may be
used to classify the operator content of the massive integrable QFT, as
explicitly shown in refs. \cite{JLG,KM,K,Smirn}.}. Finding the
general solution of eqs.\,(\ref{permu1}), (\ref{recursive}) and
(\ref{recurb}) for the IMMF poses a mathematical problem of formidable
complexity, as described in Appendix A.

Fortunately, because of an important property of the spectral
series (\ref{spectral}), an accurate knowledge of the correlation functions
can be reached with limited mathematical effort. This property consists
of a very fast rate of convergence for {\em all} distance scales
\cite{CMpol}. In view of this, the correlation functions can be determined
with remarkable accuracy by truncating the series to the first few terms
only. This statement appears to be obviously true in the infrared region
(large $M r$), where more terms included into the series only add
exponentially small contributions to the final result. The fast rate of
convergence in the crossover and ultraviolet regions seems less obvious.
In fact, for small values of the scaling variable $M r$, the
correlators usually present power-law singularities and all numbers of
particles are in principle supposed to significantly contribute to the sum.
However, this does not seem to be the case of integrable QFT with
sufficiently mild singularities in the ultraviolet region for a
``threshold suppression effect'' discussed in \cite{CMpol}. Although this
result
was originally derived for QFT with only one species of particles in the
spectrum, we expect that it also applies to the IMMF\footnote{As we will show
in Sect.\,5, this will be indeed confirmed by: (a) a direct comparison with
the numerical determination of the correlation function; (b) the
saturation of the sum-rule related to the second derivative of the
free-energy of the model
(the zero-moment of the correlation function) and (c) the saturation of the
sum-rule derived from the $c$-theorem (second moment of the correlation
function).}, due to the smooth singularity of $G(x)$ at the origin,
i.e. $G(x) \sim x^{-1/4}$.

If this crucial property of the spectral series is taken for granted,
the first terms of the series are expected to saturate the values of the
correlation function with a high degree of precision and we can only
concentrate on their analytic determination. The number of terms to be
included in the series essentially depends on the accuracy we would like to
reach in the ultraviolet region and, to this aim, it is convenient to order
them according to the energy of the particle states. For the IMMF, the first
seventeen states are collected in Table 2. The most important contributions
to the sum come from the one-particle states $A_1$, $A_2$ and $A_3$, and for
the correlation function we have correspondingly
\EQ
G(r) \,=\,\left|<0|\sigma(0)|0>\right|^2 +
\frac{\left|\Upsilon_1\right|^2}{\pi} K_0(m_1 r) +
\frac{\left|\Upsilon_2\right|^2}{\pi} K_0(m_2 r) +
\frac{\left|\Upsilon_3\right|^2}{\pi} K_0(m_3 r) +
{\cal O}\left(e^{-2 m_1 r}\right)
\label{threek}
\EN
where $K_0(x)$ is the modified Bessel function and we have defined
\EQ
\Upsilon_i \equiv <0|\sigma(0)|A_i> \,\,\,.
\label{oneparticleff}
\EN
These matrix elements will be exactly determined in next section.
Concerning the vacuum expectation value of $\sigma(x)$, it can be easily
obtained from the relationship between this field and the trace $\Theta(x)$
of the stress-energy tensor. Since $\sigma(x)$ plays the role of
the perturbing field in the theory under consideration, it
is related to $\Theta(x)$ as
\EQ
\Theta(x) \,=\, 2\pi \,h (2 - 2 \Delta_{\sigma})\,\sigma(x)
\label{theta}
\EN
On the other hand, the vacuum expectation of $\Theta(x)$ can be exactly
determined by the Thermodynamic Bethe Ansatz and its value is given
by \cite{Fateev}
\EQ
<0|\Theta(0)|0> = \frac{\pi m_1^2}{\varphi_{11}} \,\, ,
\label{vacuum}
\EN
where
\EQ
\varphi_{11} = 2 \sum_{\alpha \in {\cal A}_{11}} \sin\pi \alpha =
2\,\left(\sin\frac{2\pi}{3} + \sin\frac{2\pi}{5} + \sin\frac{\pi}{15}
\right) \,\,\,.
\EN
Hence, using the above formulas and eq. (\ref{fat}), we have
\EQ
<0|\sigma(0)|0> \,=\,\frac{4 {\cal C}^2}{15 \varphi_{11}} h^{1/15} \,=\,
(1.07496..) \,\, h^{1/15}
\label{vacuumev}
\EN

Eq.\,(\ref{threek}) provides the first terms of the large-distance expansion
of the correlation function $<\sigma(x) \sigma(0)>$. A more refined
determination of $G(x)$ may be obtained by computing the FF of the higher-mass
states, as discussed in the next section and in Appendix A.

\resection{Form Factors of the IMMF}

In the framework of the Form Factor bootstrap approach to integrable
theories, the two-particle FF play a particularly important role,
both from a theoretical and from a practical point of view.
{}From a theoretical point of view, they provide the initial conditions
which are needed for solving the recursive equations. Moreover, they
also encode all the basic properties that the matrix elements with higher
number of particles inherit by factorization, namely the asymptotic
behaviour and the analytic structure. In other words, once
the two-particle FF of the considered operator have been given, the
determination of all other matrix elements is simply reduced to solve a
well-defined mathematical problem. From a practical point of view, the
truncation of the spectral series at the two-particle level usually provides a
very accurate approximation of the correlation function, which
goes even further than the crossover region. This section is mainly devoted
to the discussion of the basic features of the two-particle FF in the IMMF.

In the general case, the FF $F^{\Phi}_{ab}(\th)$ must be a meromorphic
function of the rapidity difference defined in the strip
${\it Im}\th\in(0,2\pi)$. Its monodromy properties are dictated
by the general equations (\ref{permu1}), once specialized to the case $n=2$
\EQ
F^{\Phi}_{ab}(\th)=S_{ab}(\th)F^{\Phi}_{ab}(-\th)\,\,,
\lab{w1}
\EN
\EQ
F^{\Phi}_{ab}(i\pi+\th)=F^{\Phi}_{ab}(i\pi-\th)\,\,\,.
\lab{w2}
\EN
Thus, denoting by $F^{\it min}_{ab}(\th)$ a solution of
eqs.\,(\ref{w1}),(\ref{w2}) free of poles and zeros in the strip and
also requiring asymptotic power boundness in momenta, we conclude that
$F^{\Phi}_{ab}(\th)$ must be equal to $F^{\it min}_{ab}(\th)$ times a rational
function of $\cosh\th$. The poles of this extra function are determined
by the singularity structure of the scattering amplitude $S_{ab}(\th)$.

A simple pole in $F^{\Phi}_{ab}(\th)$ associated to the diagram of
Fig.\,8 corresponds to a positive residue simple pole in $S_{ab}(\th)$
(see eq.(\ref{spole})) and in this case we can write
\EQ
F^{\Phi}_{ab}(\th\simeq
iu_{ab}^c)\simeq\frac{i \Gamma_{ab}^c}{\th-iu_{ab}^c}F^{\Phi}_c\,\,\,.
\lab{pole}
\EN
The single particle FF $F^{\Phi}_c$ is a constant because of Lorentz
invariance.
Other poles induced by higher order singularities in the scattering amplitudes
will be considered later in this section. The kinematical poles discussed in
Section 3 do not appear at the two-particle level if the operator ${\Phi}(x)$
is local with respect to the fields which create the particles, which is the
case of interest for us.

Summarizing, the two-particle FF can be parameterized as
\EQ
F^{\Phi}_{ab}(\th)=\frac{Q^{\Phi}_{ab}(\th)}{D_{ab}(\th)}
F^{min}_{ab}(\th)\,\,,
\lab{param}
\EN
where $D_{ab}(\th)$ and
$Q^{\Phi}_{ab}(\th)$ are polynomials in $\cosh\th$: the
former is fixed by the singularity structure of $S_{ab}(\th)$ while
the latter carries the whole information about the operator ${\Phi}(x)$.

An upper bound on the asymptotic behaviour of FF and then on the order
of the polynomial $Q^{\Phi}_{ab}(\th)$ in eq.\,(\ref{param})
comes from the following argument. Let $2\Delta_{\Phi}$ be the scaling
dimension of the scalar operator ${\Phi}(x)$ in the ultraviolet limit, i.e.
\EQ
<{\Phi}(x){\Phi}(0)>\,\sim\frac{1}{\,\,\,|x|^{4\Delta_{\Phi}}}\,\,,
\hs\hs\hs\hs|x|\goto 0\,\,.
\EN
Then in a massive theory
\EQ
M_p\equiv\int d^2x\,|x|^p<{\Phi}(x){\Phi}(0)>_c
\hspace{.5cm} < +\infty \hspace{1cm} {\rm if} \hspace{.8cm}
p > 4\Delta_{\Phi}-2\,\,,
\lab{moment}
\EN
where the subscript $c$ denotes the connected correlator.
The two-point correlator may be expressed in terms of its
euclidean Lehmann representation as
\EQ
<\Phi(x)\Phi(0)>_c=\int d^2p\,\,e^{ipx}\int
d\mu^2\,\,\frac{\rho(\mu^2)}{p^2+\mu^2}\,\,\,,
\label{leh}
\EN
where the spectral function $\rho$ is given by
\[
\rho(\mu^2)=\frac{1}{2\pi}\sum_{n=1}^{\infty}\int_{\th_1>\ldots>\th_n}
\frac{d\th_1}{2\pi}\ldots\frac{d\th_n}{2\pi}|F^{\Phi}_{a_1,\ldots,a_n}
(\th_1,\ldots,\th_n)|^2\delta(\sum_{k=1}^{n}m_k\cosh\th_k-\mu)
\delta(\sum_{k=1}^nm_k\sinh\th_k)\,\,\,,
\]
Substituting eq.\,(\ref{leh}) into the definition of $M_p$ and performing
the integrations over $p$, $\mu$, and $x$, one finds
\EQ
M_p\sim\sum_{n=1}^\infty\int_{\th_1>\ldots>\th_n}\,\,
d\th_1\ldots d\th_n \frac{|F^{\Phi}_{a_1,\ldots,a_n}
(\th_1,\ldots,\th_n)|^2} {\left(\sum_{k=1}^n m_k\cosh\th_k\right)^{p+1}}\,\,
\delta\left(\sum_{k=1}^nm_k\sinh\th_k\right)\,\,\,.
\lab{aaa}
\EN
Eq.\,(\ref{moment}) can now be used to derive an upper bound for the
real quantity $y_{\Phi}$, defined by
\EQ
\lim_{|\th_i|\goto\infty}
F^{\Phi}_{a_1,\ldots,a_n}(\th_1,\ldots,\th_n) \sim \,
e^{y_{\Phi}|\th_i|}\,\,\,.
\lab{bound}
\EN
This can be achieved by firstly noting that taking the limit
$\th_i\goto+\infty$ in the integrand of eq.\,(\ref{aaa}), the delta-function
forces some other rapidity $\th_j$ to diverge to minus infinity as
$-\th_i$. Since the matrix element
$F^\Phi_{a_1,\ldots,a_n}(\th_1,\ldots,\th_n)$ depends on the rapidity
differences, it will contribute to the integrand a factor
$e^{2y_\Phi|\th_i|}$ in the limit $|\th_i|\goto\infty$.
Then eq.\,(\ref{moment}) leads to the constraint
\EQ
y_{\Phi}\,\leq\,\Delta_\Phi\,\,\,.
\lab{bbb}
\EN
Note that this conclusion may not hold for non-unitary theories
since not all the terms in the expansion over intermediate states are
guaranteed to be positive in these cases.

Let us see how the aforementioned considerations apply to the specific
case of the IMMF. An appropriate solution of eqs.\,(\ref{w1}) and
(\ref{w2}), corresponding to the scattering amplitudes reported in Table 1,
can be written as
\EQ
F^{min}_{ab}(\th)=\left(-i\sinh\frac{\th}{2}\right)^{\delta_{ab}}
\prod_{\alpha\in{\cal A}_{ab}}\left(G_{\alpha}(\th)
\right)^{p_\alpha}\,\,,
\lab{fmin}
\EN
where
\EQ
G_{\al}(\th)=\exp\left\{2\int_0^\infty\frac{dt}{t}\frac{\cosh\left(
\al - \frac{1}{2}\right)t}{\cosh\frac{t}{2}\sinh
t}\sin^2\frac{(i\pi-\th)t}{2\pi}\right\}\,\,\,.
\lab{block}
\EN
For large values of the rapidity
\EQ
G_{\al} (\th) \,\sim\, \exp(|\th|/2) \,\,\, , |\th|\goto\infty \,\,,
\label{dec}
\EN
independent of the index $\al$. Other properties of this function are
discussed in Appendix B.

An analysis of the two-particle FF singularities which will be described
later on in this section, suggests that the pole terms appearing
in the general parameterization eq.\,(\ref{param}) could be written as
\EQ
D_{ab}(\th)=\prod_{\alpha\in {\cal
A}_{ab}} \left({\cal P}_\alpha(\th)\right)^{i_\alpha}
\left({\cal P}_{1-\alpha}(\th)\right)^{j_\alpha} \,\,\,,
\lab{dab}
\EN
where
\EQ
\begin{array}{lll}
i_{\alpha} = n+1\,\,\, , & j_{\alpha} = n \,\,\, , &
{\rm if} \hspace{.5cm} p_\alpha=2n+1\,\,\,; \\
i_{\alpha} = n \,\,\, , & j_{\alpha} = n \,\,\, , &
{\rm if} \hspace{.5cm} p_\alpha=2n\,\,\, ,
\end{array}
\EN
and we have introduced the notation
\EQ
{\cal P}_{\al}(\th)\equiv
\frac{\cos\pi\al-\cosh\th}{2\cos^2\frac{\pi\al}{2}}\,\,\,.
\lab{polo}
\EN
Both $F^{min}_{ab}(\th)$ and $D_{ab}(\th)$ have been normalized
to 1 in $\th=i\pi$.

Finally, let us turn our attention to the determination of the
polynomials $Q^{\Phi}_{ab}(\th)$ for the specific operator we are
interested in, namely the magnetization field $\sigma(x)$. In view of the
relation (\ref{theta}), this is the same as considering the analogous problem
for the trace of the energy-momentum tensor $\Theta(x)$. For reasons which
will become immediately clear, we will consider the latter operator in
the remainder of this section.

The conservation equation $\partial_\mu T^{\mu\nu}=0$ implies the following
relations among the FF of the different components of the energy-momentum
tensor
\bea
F^{T^{++}}_{a_1,\ldots,a_n}(\th_1,\ldots,\th_n) &\sim& \frac{P^+}{P_-}
F^{\Theta}_{a_1,\ldots,a_n}(\th_1,\ldots,\th_n)\,\,\,;\\
F^{T^{--}}_{a_1,\ldots,a_n}(\th_1,\ldots,\th_n) &\sim& \frac{P^-}{P_+}
F^{\Theta}_{a_1,\ldots,a_n}(\th_1,\ldots,\th_n)\,\,,
\lab{mismatch}
\eea
where $x^\pm=x^0\pm x^1$ are the light-cone coordinates
and $P^\pm\equiv\sum_{i=1}^n p_{a_i}^\pm$. The requirement that all
the components of the energy-momentum tensor must exhibit the same
singularity structure, leads to conclude that the FF of $\Theta(x)$
must contain a factor $P^+P^-$. However, the case $n=2$ is special because, if
the two particles have equal masses, the mismatch of the singularities
disappears in eqs.\,(\ref{mismatch}) and no factorisation takes place. From
this analysis, we conclude that for our model we can write
\EQ
Q^\Theta_{ab}(\th) = \left(\cosh\th +
\frac{m_a^2+m_b^2}{2m_am_b}\right)^{1-\delta_{ab}} P_{ab}(\th)\,\,,
\EN
where
\EQ
P_{ab}(\th)\equiv\sum_{k=0}^{N_{ab}} a^k_{ab}\,\cosh^k\th\,\,\,.
\EN
The degree $N_{ab}$ of the polynomials $P_{ab}(\th)$ can be severely
constrained by using eqs.\,(\ref{bound}) and (\ref{bbb}). Additional
conditions for these polynomials are provided by the normalization
of the operator $\Theta(x)$, that for the diagonal elements
$F^\Theta_{aa}$, reads
\EQ
F_{aa}^\Theta(i\pi)=<A_a(\th_a)|\Theta(0)|A_a(\th_a)>=2\pi m^2_a\,\,\,.
\label{ipi}
\EN
Using all the information above, we can now proceed in the computation of
the IMMF form factors, starting from the simplest two-particle FF of the
model, namely $F_{11}^{\Theta}(\th)$ and $F_{12}^{\Theta}(\th)$.

First of all, by using eqs.\,(\ref{bbb}) and (\ref{dec}) one concludes
that $N_{11}\leq 1$ and
$N_{12}\leq 1$. In view of the normalization condition (\ref{ipi}), only one
unknown parameter, say $a^1_{11}$, is necessary in order to have the complete
expression of $F^\Theta_{11}(\th)$. On the contrary, we need two parameters,
$a^0_{12}$ and $a^1_{12}$, to specify $F^\Theta_{12}(\th)$. To determine
all three unknown parameters, note that the scattering
amplitude $S_{11}(\th)$ possesses three positive residue poles at
$\th = i\frac{2\pi}{3}$,
$\th = i\frac{2\pi}{5}$ and $\th = i\frac{\pi}{15}$ which correspond to the
particles $A_1$, $A_2$ and $A_3$ respectively; on the other hand,
$S_{12}(\th)$ exhibits four positive residue poles at
$\th = i\frac{4\pi}{5}$, $\th = i\frac{3\pi}{5}$, $\th = i\frac{7\pi}{15}$ and
$\th = i\frac{4\pi}{15}$ associated to $A_1$, $A_2$, $A_3$ and $A_4$.
Hence, since three poles are common to both amplitudes and no multiple poles
appear in both of them, eq.\,(\ref{pole}) provides a system of three linear
equations which uniquely determine the coefficients $a^1_{11}$, $a^0_{12}$
and $a^1_{12}$
\EQ
\frac{1}{\Gamma_{11}^c} {\rm Res}_{\th=iu_{11}^c}F_{11}^{\Theta}(\th)=
\frac{1}{\Gamma_{12}^c} {\rm Res}_{\th=iu_{12}^c}F_{12}^{\Theta}(\th)
\hspace{1cm}c=1,2,3\,\,.
\EN
The result of this calculation can be expressed in terms of the mass
ratios $\hat m_i = m_i/m_1$ as
\EQ
P_{11}(\th) \,=\,\frac{2\pi m_1^2}{\hat m_3 \hat m_7} (2\cosh\th +
2 + \hat m_3 \hat m_7)
\EN
\EQ
P_{12}(\th) \,=\,H_{12} \left(2 \hat m_2 \cosh\th + \hat m_2^2 +
\hat m_8^2\right)
\EN
where
\[
H_{12} = (1.912618..) \, m_1^2 \,\,\,.
\]
Equations (\ref{pole}) can now be used to obtain the one-particle
form factors $F^{\Theta}_a$ ($a=1,\ldots,4$), whose numerical values are
reported in Table 3. In particular,
\[
F^\Theta_1 \,=\, \frac{\pi m^2_1}{\Gamma_{11}^1 \hat m_3 \hat m_7}
\frac{(1 + \hat m_3 \hat m_7)\,
\left[G_{\frac{2}{3}} G_{\frac{2}{5}} G_{\frac{1}{15}}
\left(\frac{2 \pi i}{3}\right) \right]\cos^2\frac{\pi}{5} \cos^2\frac{\pi}{30}
}
{\sin\frac{8\pi}{15} \,\sin\frac{2\pi}{15} \,\sin\frac{3\pi}{10}\,
\sin\frac{11\pi}{30}}
\]
\[
F^{\Theta}_2 \,=\, -\frac{4 \pi m_1^2}{\Gamma_{11}^2 \hat m_3 \hat m_7}
\frac{\sin\frac{\pi}{5}}{\sin\frac{2 \pi}{5}}
\frac{ \left(2 \cos\frac{2\pi}{5} + 2 + \hat m_3 \hat m_7\right)\,
\left[G_{\frac{2}{3}} G_{\frac{2}{5}} G_{\frac{1}{15}}
\left(\frac{2\pi i}{5}\right)\right]
\left(\cos\frac{\pi}{3}\,\cos\frac{\pi}{30}\,\cos\frac{\pi}{5}\right)^2
}
{\sin\frac{2\pi}{5} \,\sin\frac{8\pi}{15}\,
\sin\frac{\pi}{6} \,\sin\frac{7\pi}{30}}
\]
\[
F^{\Theta}_3 \,=\, \frac{4 \pi m_1^2}{\Gamma_{11}^3 \hat m_3 \hat m_7}
\frac{\sin\frac{\pi}{30}}{\sin\frac{\pi}{15}}
\frac{\left(2 \cos\frac{\pi}{15} + 2 + \hat m_3 \hat m_7\right)\,
\left[G_{\frac{2}{3}} G_{\frac{2}{5}} G_{\frac{1}{15}}
\left(\frac{i\pi}{15}\right) \right]
\left(\cos\frac{\pi}{5} \cos\frac{\pi}{3} \cos^2\frac{\pi}{30}\right)^2
}
{\sin\frac{3\pi}{10}\,\sin\frac{11\pi}{30}\,
\sin\frac{\pi}{6}\,\sin\frac{7\pi}{30}}
\]

In order to continue in the bootstrap procedure and compute the other
one-particle and two-particle FF, we have to firstly consider the
multiple poles of the scattering amplitudes. Such poles are known to
represent the two-dimensional analog of anomalous thresholds associated to
multi-scattering processes \cite{ColT}. These are processes in which the two
ingoing particles decay into their ``constituents'', which interact and
then recombine to give a two-particle final state. In the general framework of
relativistic scattering theory, the location of this kind of singularities
is determined by the so-called Landau rules \cite{eden}. In the
two-dimensional case, such rules admit the following simple formulation:
singularities occur only for those values of the momenta for which a
space-time diagram of the process can be drawn as a geometrical figure with
all (internal and external) particles on mass-shell and energy-momentum
conservation at the vertices. The simplified two-dimensional kinematics
only selects discrete values of the external momenta for which such a
construction is possible and this is the reason why in two dimensions the
``anomalous'' singularities appear as poles rather than branch cuts. The
order of the pole and its residue can be determined using the Cutkosky rule
\cite{eden} which states that the discontinuity across the singularity
associated to the abovementioned diagram is obtained evaluating it as if it
were a Feynman graph but by inserting the complete scattering amplitudes at the
interaction points and by replacing the internal propagators with
mass-shell delta-functions $\theta(p^0)\,\delta(p^2-m^2)$. For a diagram
containing $P$ propagators and $L$ loops, $P-2L$ delta-functions
survive the $L$ two-dimensional integrations; since the singularity
whose discontinuity is a single delta-function is a simple pole, the
graph under consideration leads to a pole of order $P-2L$ in the
amplitude \cite{BCDS}.

Let us initially consider the second order poles. A second order pole at
$\th=i\varphi$ occurs in the amplitude $S_{ab}(\th)$ if one of
the two diagrams in Figures 9.a and 9.b can be actually drawn, namely if
\EQ
\eta \equiv \,\pi-u_{cd}^a-u_{de}^b \in[0,\pi)\,\,\, .
\lab{constraint}
\EN
The quantity $i\eta$ is the rapidity difference between the intermediate
propagating particles $A_c$ and $A_e$. From these figures, it is easy to
see that
\EQ
\varphi=u_{ad}^c+u_{db}^e-\pi\,\,\,.
\EN
The crossing symmetry expressed by eq.\,(\ref{crossing}) obviously implies
that, in addition to the double pole in $\th=i\varphi$, an analogous pole must
be present in $\th=i(\pi-\varphi)$. Since the residues of the two
poles are now equal, it is impossible to distinguish between a direct
and a crossed channel, and the two poles must be treated on exactly
the same footing. At the diagrammatic level, this fact is reflected by
the possibility to find a diagram satisfying eq.\,(\ref{constraint})
also for $\th=i(\pi-\varphi)$. Hence, let us consider only
one of these poles, the one located at $\th=i\varphi$. In the vicinity
of this pole, the scattering amplitude can be approximated as (see Fig.\,9.a)
\EQ
S_{ab}(\th) \,\simeq\,
\frac{(\Gamma_{cd}^a \Gamma_{de}^b)^2S_{ce}(\eta)}{(\th-i\varphi)^2}\,\,\,.
\lab{res2}
\EN
Note that the expression of this residue, which is obtained for
$\eta>0$, is also valid in the limiting situation $\eta=0$ (Fig.\,9.b), for
which a residue $(\Gamma_{cd}^a \Gamma_{de}^b \Gamma_{cf}^b \Gamma_{fe}^a)$ is
expected. In fact, the consistency of the theory requires
\EQ
\Gamma_{cd}^a \Gamma_{de}^b \, =\, \Gamma_{cf}^b \Gamma_{fe}^a\,\, ,
\lab{identity}
\EN
an equation that is indeed satisfied for the three-point couplings of the IMMF.
Moreover, in the case $\eta=0$, the ``fermionic'' nature of the particles,
expressed by the relations
\EQ
S_{ab}(0) \,=\,
\left\{
\begin{array}{cl}
-1 & if\hspace{.5cm} a=b \,\,\, ; \\
1  & if\hspace{.5cm} a\neq b\,\,\, ,
\end{array}
\right.
\EN
implies that the two particles $A_c$ and $A_e$ propagating with the
same momentum in Fig.\,9.b cannot be of the same species. In this case, the
factor $S_{ce}(\eta=0)$ in eq.(\ref{res2}) equals unity.

The double pole at $\th=i\varphi$ in $S_{ab}(\th)$ induces a singularity
at the same position in $F_{ab}^{\Phi}(\th)$. For $\eta > 0$,
this is associated to the diagram on the left hand side of Fig.\,10. Since
the singularity is now determined by a single triangular loop, the form
factor $F_{ab}^{\Phi}(\th)$ has only a simple pole rather than a double pole.
The residue is given by
\EQ
\Gamma_{cd}^a \Gamma_{de}^b S_{ce}(\eta) F_{ce}^{\Phi}(-\eta)\,\,\,.
\EN
Eq.\,(\ref{w1}) can now be used to write (see the right hand side of
Fig.\,10)
\EQ
F_{ab}^{\Phi}(\th\simeq i\varphi) \simeq
i \,\frac{\Gamma_{cd}^a \Gamma_{de}^b F_{ce}^{\Phi}(\eta)}{\th-i\varphi}\,\,\,.
\lab{ffres2}
\EN
As written, this result also holds for $\eta=0$.

The poles of order $p>2$ in the scattering amplitudes and the
corresponding singularities in the two-particle FF can be treated
as a ``composition'' of the cases $p=1$ and $p=2$. This is the case, for
instance, of a third order pole with positive residue at $\th=i\varphi$ in
$S_{ab}(\th)$. In the $S$-matrix, a third order pole occurs if the scattering
angle $\eta$ in Fig.\,9.a coincides with the resonance angle $u_{ce}^f$.
The corresponding diagram is drawn in Figure 11 and in this case we have
\EQ
S_{ab}(\th\simeq i\varphi)\simeq i \,
\frac{(\Gamma_{cd}^a \Gamma_{de}^b \Gamma_{ec}^f)^2}{(\th-i\varphi)^3}\,\,\,.
\lab{res3}
\EN
The third-order pole at the crossing-symmetric position $\th=i(\pi-\varphi)$
has negative residue since it corresponds to the crossed channel pole. With
respect to the case $p=2$, the pole at $\th=i\varphi$ in $F_{ab}^{\Phi}(\th)$
becomes double (see Figure 12)
\EQ
F_{ab}^{\Phi}(\th\simeq i\varphi)\simeq -\,
\frac{\Gamma_{cd}^a \Gamma_{de}^b \Gamma_{ec}^f}{(\th-i\varphi)^2}
F_f^{\Phi}\,\,\,,
\lab{ffres3}
\EN
while the pole at $\th=i(\pi-\varphi)$ stays simple.

The above analysis suggests the validity of the following general
pattern for the structure of the form-factor poles: a pole
of order $2n$ at $\th=i\varphi$ in the crossing-symmetric scattering amplitude
$S_{ab}(\th)$ will induce a pole of order $n$ both at $\th=i\varphi$ and
at $\th=i(\pi-\varphi)$ in the two-particle form factor $F_{ab}^{\Phi}(\th)$;
viceversa, a positive residue pole of order $(2n+1)$ at $\th =
i\varphi$ in $S_{ab}(\th)$ will induce a pole of order $(n+1)$ at
$\th = i \varphi$ and a pole of order $n$ at the crossing symmetrical
position $\th=i(\pi-\varphi)$ in $F_{ab}^{\Phi}(\th)$.
We have used this result to write the parameterization
of eq.\,(\ref{dab}).

Moreover, in integrable QFT, these arguments can be easily extended to the
higher matrix elements $F_{a_1,\ldots,a_n}^\Phi(\th_1,\ldots,\th_n)$ with
$n>2$, since the complete factorisation of multiparticle processes
prevents the generation of new singularities. In other words, the singularity
structure of the n-particle FF is completely determined by the
product of the poles present into each two-particle sub-channel.

We have used eqs.\,(\ref{pole}), (\ref{ffres2}), (\ref{ffres3}) to continue
the bootstrap procedure for the two-particle FF of $\Theta(x)$ in the IMMF
up to the level $A_3A_3$, and an illustration of the method through a
specific example may be found in Appendix C. The results obtained for the
coefficients $a^k_{ab}$ (the only unknown quantities in the parameterization
of eq.\,(\ref{param}) after the pole structure has been fixed) are
summarized for convenience in Table 4; Table 3 contains the complete list of
the
one-particle matrix elements. Two important comments are in order here. The
first is that in all the determinations of $F_{ab}^\Theta$
except $F_{11}^\Theta$ and $F_{12}^\Theta$, a number of equations larger
than the number of unknown
parameters is obtained. The fact that these overdetermined systems of
equations always admit a solution\footnote{Solutions can only be found by
choosing the three-point couplings $\Gamma_{ab}^c$ either all positive
or all negative. Hence, this restricts the ambiguity of the three-point
couplings to an overall $\pm $ sign only. We are not aware of
any other explanation for this constraint on the $\Gamma_{ab}^c$.}
provides a highly nontrivial check of the
results of this section.
The second point is that, since the pole structure has been identified, there
is no obstacle, in principle, to continue further the bootstrap procedure
and to achieve any desired precision in the determination of the correlation
function in the ultraviolet region. Actually, we will show in the next section
that the information contained in Tables 3 and 4 are more than enough for
practical purposes. Nevertheless, from a purely theoretical point of view, it
would be obviously desirable to have a complete solution of the recursive
equations. A possible approach to this non trivial mathematical problem is
suggested in Appendix A.

\resection{Comparison with Numerical Simulations}

The data collected in Tables 3 and 4, together with the vacuum
expectation value eq.\,(\ref{vacuum}) and the three-particle matrix
element $F_{111}^\Theta(\th_1,\th_2,\th_3)$ given in Appendix A
provide us with the complete large-distance expansion of the
correlator $<\Theta(x)\Theta(0)>$ up to order $e^{-(m2+m3)|x|}$. A
first check of the degree of convergence of the series, and then of its
practical utility, is obtained by exploiting the exact knowledge of the
second and zeroth moments of the correlation function we are considering.
Indeed, in a massive theory the $c$-theorem sum rule provides the
relation \cite{Zamcth}
\EQ
C = \frac{3}{4\pi}\int d^2x|x|^2<\Theta(x)\Theta(0)>_c\,\,\,,
\lab{cth}
\EN
where $C$ is the central charge of the conformal theory describing the
ultraviolet fixed point. For the Ising model $C=\frac{1}{2}$.
In addition, if we write the singular part of the free energy
per unit volume as $f_s\simeq-UM^2(h)$, a double differentiation with
respect to $h$ leads to the identity
\EQ
U = \frac{1}{\pi^2}\int d^2x<\Theta(x)\Theta(0)>_c\,\,\,.
\lab{bulk}
\EN
On the other hand, the exact value of the universal amplitude $U$ is
obtained by plugging eq.\,(\ref{vacuum}) into
\EQ
U = \frac{4\pi}{M^2(h)}<0|\Theta|0> = 0.0617286..
\EN
The contributions to the sum rules (\ref{cth}) and (\ref{bulk})
from the first eight states in the spectral representation of
the connected correlator are listed in the Tables 5 and 6, together
with their partial sums. The numerical data are remarkably close to their
theoretical values. Notice that a very fast saturation is also observed
in the case of the zeroth moment, despite the absence of any suppression
of the ultraviolet singularity.

Let us now directly compare the theoretical prediction of the connected
correlation function $G_c(x) = <\sigma(x) \sigma(0)>_c$ with its numerical
evaluation. A collection of high-precision numerical estimates of $G_c(x)$,
for different values of the magnetic field $h$ and different size
$L$ of the lattices, can be found in the reference \cite{LR}. We have decided
to consider the set of data relative to $L=64$ and $h=0.075$, where
the numerical values of $G(x)$ are known on $32$ lattice space (Table 7).
Such a choice was dictated purely by the requirement to use data where the
effects of numerical errors are presumed to be minimized. Errors can be in
fact induced either from the finite size $L$ of the sample or from the
residue fluctuations of the critical point, which may be not sufficiently
suppressed for small values of the magnetic field $h$.

In order to compare the numerical data with our theoretical determination,
we only need to fix two quantities. The first consists in extracting the
relationship between the inverse correlation length, expressed in lattice
units, and the mass scale $M(h)$ entering the form factor expansion. The
second quantity we need is the relative normalization of the operator
$\sigma(x)$ defined on the lattice, denoted by $\sigma_{\rm lat}(x)$, with
the operator $\sigma(x)$ entering our theoretical calculation in the
continuum limit. Let us consider the two issues separately.

The correlation length $\xi$ is easily extracted by using eq.\,(\ref{threek})
to analyse the exponential decay of the numerical data
collected in Table 7. As a best fit of this quantity, we obtain
\EQ
M(h =0.075) \,=\,\xi^{-1} \,=\,5.4(3)\,\,\,.
\label{latcor}
\EN

Let us turn our attention to the second problem. The easiest way to set the
normalization of $\sigma(x)$ with respect to $\sigma_{\rm lat}(x)$ is to
compare their vacuum expectation value. The lattice determination of this
quantity can be found in \cite{D,LR} and, within the numerical precision,
it is given by
\EQ
<\sigma_{\rm lat}(0)> \,=\,1.000(1) \,h^{1/15} \,\,\, .
\EN
On the other hand, the theoretical estimate of $<\sigma(0)>$ was given in
eq. (\ref{vacuumev}). Hence, comparing the two results, the relative
normalization is expressed by the constant ${\cal N}$ as
\EQ
\sigma_{\rm lat}(x) \,=\,{\cal N}\,\sigma(x) \,=\,0.930(3)\, \,\sigma(x)
\,\,\, .
\label{latcont}
\EN

Once these two quantities are fixed, there are no more adjustable
parameters to compare the numerical data with the large-distance expansion
of $G_c(x)$. The form factors of the field $\sigma(x)$, entering the series
(\ref{spectral}) can be easily recovered from those of $\Theta(x)$, by using
the relationship of these fields given by (\ref{theta}), and for the
correlation function we have
\EQ
<\sigma(x) \sigma(0)>_c \,=\,\left(\frac{4}{15 \pi h}\right)^2
\,<\Theta(x) \Theta(0)>_c\,\,\,.
\EN
The comparison between the two determinations of
$G_c(x)$ can be found in Figures 13 and 14. In Fig.\,13 we have only
included the first three terms of $G_c(x)$ (those relative to the form factors
of the one-particle states $A_1$, $A_2$, $A_3$). As shown in this
figure, they can reproduce correctly the behaviour of the
correlation function on the whole infrared and crossover regions. A slight
deviation of the theoretical curve from the numerical values is only observed
for the first points of the ultraviolet region, where a better approximation
can be obtained by including more terms in the form factor series. This is
shown in Figure 14, where five more contributions (those relative
to form factors up to state $A_1 A_3$) have been added to the series.

\resection{Conclusion}

The basic results of this paper can be summarised as follows. The
Zamolodchikov S-matrix for the IMMF has been used as the starting
point to implement a bootstrap program for the FF of the magnetization
operator. Although the general solution of the bootstrap recursive
equations remains a challenging mathematical problem, the matrix elements
yielding the main contributions to the spectral representation of the
correlator
$G(x)=<\sigma(x)\sigma(0)>$ have been explicitly
computed. This has enabled us to write a large-distance expansion for
$G(x)$ which is characterised by a very fast rate of convergence and
provides accurate theoretical predictions for comparison with data
coming from high precision numerical simulations.

It would be interesting to obtain analogous results for the other
relevant operator of the theory, namely the energy density
$\varepsilon(x)$. To this aim, the only difficulty one has to face is
the determination of the initial conditions for the form factor
bootstrap equations appropriate for this operator.
In the case of the field $\sigma(x)$,
we solved this problem by exploiting the proportionality with the trace
$\Theta(x)$ of the energy-momentum tensor. Notice that, due to the absence of
symmetries in the space of states of the IMMF, the occurrence of the
polynomials $Q_{ab}^\Phi(\th)$ in the two-particle FF is precisely what is
needed in order to distinguish between the matrix elements of
$\sigma(x)$ and those of $\varepsilon(x)$.

In conclusion, it must be remarked that the methods discussed in this paper
can be generally used within the framework of integrable
QFT. As a matter of fact, here they have been applied to a model which,
for the absence of internal symmetries and the richness of its
pole structure, can be considered as an extreme case of complexity.
For instance, similar results
to those contained in this paper can be obtained for other
physically interesting situations, such as the thermal deformations of the
tricritical Ising and three-state Potts models. The exact S-matrices
for these models were determined in ref.\,\cite{ChM,Zamo-Fateev}.

\vspace{1cm}
{\em Acknowledgments}. We are grateful to J.L. Cardy for useful discussions.

\newpage

\appendix

\appsection

Aim of this appendix is to formulate in general terms the mathematical
problem related to the computation of a generic form factor of the
scalar operators in the IMMF.
Such a formulation is obtained by exploiting a decisive property of the
model, namely its bootstrap structure: {\em any} particle $A_i$
($i=1,2,\ldots 8$) of the theory appears as a bound state of some scattering
process involving the fundamental particle $A_1$ and therefore can be obtained
by a sufficient number of fusions of the particles $A_1$'s alone. The
simplest examples are provided by the particles $A_2$ and $A_3$, which appear
in the initial amplitude $S_{11}(\theta)$. Hence, the bootstrap structure of
the model implies that all possible FF of the theory can be in principle
obtained by the $n$-particle FF which only involve the particle $A_1$, by
simply applying the residue equations (\ref{recurb}) the number of times we
need to reach the FF under consideration. For instance, the form factors
$F_{22}(\th)$ and $F_{33}(\th)$ can be both obtained by starting
from $F_{1111}$ and by applying twice (\ref{recurb}) on the poles
at $\frac{2 \pi i}{5}$ and $\frac{i \pi}{15}$, respectively.

In view of the role played by the FF with the particles $A_1$, it is
convenient to use a convenient notation. For brevity, we will denote them as
$F_n(\th_1,\th_2,\ldots,\th_n)
\equiv F_{11\ldots 1}(\th_1,\ldots,\th_n)$. It is now quite easy to
find a parameterization of $F_n$ which correctly takes into account their
monodromy properties and the pole structure. It can be written as
\EQ
F_n(\th_1,\ldots,\th_n) = H_n \frac{\Lambda_n(x_1,\ldots,x_n)}
{(\omega_n(x_1,\ldots,x_n))^n} \prod_{i<j}
\frac{F_{11}^{min}(\th_{ij})}{D_{11}(\th_{ij}) (x_i+x_j)} \,\,\, .
\label{parame}
\EN
Let us explain the origin of each term entering the above equation.

The monodromy equations (\ref{permu1}) can be satisfied in terms of the
functions $F_{11}^{min}(\th)$, solution of the equations
\EQ
\begin{array}{ccc}
F_{11}^{min}(\th) &=& S_{11}(\th)\, F_{11}^{min}(-\th)\,\, , \\
F_{11}^{min}(i\pi-\th) &=& F_{11}^{min}(i\pi + \th) \,\, .
\end{array}
\label{permu2}
\EN
These functions are required to have neither zeros or poles in the strip
$(0, 2 \pi i)$. $F_{11}^{min}(\th)$ can be explicitly written in terms of the
functions $G_{\lambda}(\th)$ discussed in the appendix B as
\EQ
F_{11}^{min}(\th) = -i \sinh\frac{\th}{2} \,G_{\frac{2}{3}}(\th) \,
G_{\frac{2}{5}}(\th) \,G_{\frac{1}{15}}(\th) \,\,\, .
\EN
Once the monodromy properties of $F_n$ are taken into account,
we have to consider their pole structure. Note that, apart from the
product of the $F_{11}^{min}(\th_{ij})$'s, the remaining
part of these amplitudes can only be expressed in terms of functions
of the variables $\th_{ij}$ which are even and $2 \pi i$ periodic, i.e.
functions of the variables $\cosh\th_{ij}$. Equivalently, they have
to be symmetric functions of the variables $x_i\equiv e^{\th_i}$.
A basis in the space of the symmetric functions of $n$-variables is provided
by the elementary symmetric polynomials $\omega_i(x_1,x_2,\ldots,x_n)$
\cite{sym}, defined by the generating function
\EQ
\prod_{k=1}^n (x+x_i) = \sum_{j=0}^n x^{n-j} \omega_j(x_1,\ldots,x_n) \,\,\,.
\EN
The bound state poles of $F_n$ in all possible subchannels of
the amplitude $F_n$ is encoded in the product of the terms
\EQ
D_{11}(\th) \equiv
{\cal P}_{\frac{2}{3}}(\th_{ij}) \, {\cal P}_{\frac{2}{5}}(\th_{ij})
{\cal P}_{\frac{1}{15}}(\th_{ij})\,\,\,,
\EN
where ${\cal P}_{\lambda}(\theta)$ is defined in eq.\,(\ref{polo}).
Concerning the kinematical poles, all of them are present in the product
$\prod_{i<j} (x_i + x_j)$. Finally, in (\ref{parame}) $H_n$ is a
normalization constant, $\Lambda_n(x_1,x_2,\ldots,x_n)$ is a symmetric
polynomial and the last term $(\omega_n(x_1,\ldots,x_n))^n$ (which has no
zeros in the physical strip) has been inserted in order to have a convenient
form of the recursive equations.

The polynomials $\Lambda_n(x_1,x_2,\ldots,x_n)$ can be obtained by solving
the recursive
equations (\ref{recursive}) and (\ref{recurb}). Using the parameterization
(\ref{parame}), for the bound-state recursive equations we have
\EQ
\frac{\Lambda_{n+1}(e^{i \pi/3} x, e^{-i\pi/3} x,x_2,\ldots,x_n)}
{x^{n+3} \left[\prod_{k=1}^n (x+x_k) (x-\eta x_k) (x-\eta^{-1} x_k)\right]}
 \,=\,
\Lambda_n(x,x_2,\ldots,x_n) \,\, ,
\label{pinc2}
\EN
where $\eta = \exp(i \pi/15)$. In writing (\ref{pinc2}) we have
choose the normalization constants $H_n$ as
\EQ
H_{n+1} = \,
\frac{\Gamma_{11}^1 \sin\frac{2\pi}{15} \sin\frac{11\pi}{30}
\sin\frac{8\pi}{15} \sin\frac{3\pi}{10}}
{2 G_{11}\left(\frac{2 \pi i}{3}\right)
\left(\cos \frac{\pi}{3} \cos\frac{\pi}{5} \cos\frac{\pi}{30}\right)^2}
\left[\frac{\sin^2\frac{11\pi}{30}}{4 \gamma
\left(\cos \frac{\pi}{3} \cos\frac{\pi}{5} \cos\frac{\pi}{30}\right)^2}
\right]^n
\,\, H_n \,\,\,.
\EN
and, for simplicity, we have introduced the notation
$G_{11}(\th) \equiv G_{\frac{2}{3}}(\th) \,
G_{\frac{2}{5}}(\th) \,G_{\frac{1}{15}}(\th)$. The constant $\gamma$ is
given by
\EQ
\gamma = \frac{G_{\frac{2}{3}}\left(0\right)
G_{\frac{2}{3}}\left(\frac{2\pi i}{3}\right)
G_{\frac{2}{5}}\left(-\frac{i\pi}{3}\right)
G_{\frac{2}{5}}\left(\frac{i\pi}{3}\right)
G_{\frac{1}{15}}\left(-\frac{i\pi}{3}\right)
G_{\frac{1}{15}}\left(\frac{i\pi}{3}\right)}
{G_1 \left(\frac{i\pi}{3}\right)
G_{\frac{2}{3}}\left(\frac{i\pi}{3}\right)
G_{\frac{11}{15}}\left(0\right)
G_{\frac{1}{15}}\left(0\right)
G_{\frac{2}{5}}\left(0\right)
G_{-\frac{4}{15}}\left(0\right)} \,\,\,.
\EN
A second hierarchy of recursive equations is obtained from the residue
conditions of the kinematical poles, eq.\,(\ref{recursive}), which
can be written as
\EQ
(-)^n \Lambda_{n+2}(-x,x,x_1,\ldots,x_n) \,=\, {\cal A}_n\, U(x,x_1,\ldots,x_n)
\Lambda_n(x_1,\ldots,x_n)
\,\, ,
\EN
where
\EQ
{\cal A}_n = \,
\frac{\gamma G_{11}^2\left(\frac{2\pi i}{3}\right)
\left(2 \cos\frac{\pi}{3} \cos\frac{\pi}{5} \cos\frac{\pi}{30}\right)^6}
{\sin^2\frac{11\pi}{30} \left(\Gamma_{11}^1
\sin\frac{2\pi}{15} \sin\frac{11\pi}{30}
\sin\frac{8\pi}{15} \sin\frac{3\pi}{10}\right)^2}
\left[
\frac{\gamma^2 \sin\frac{2\pi}{3} \sin\frac{2\pi}{5} \sin\frac{\pi}{15}}
{8 G_{11}(0) \sin^4\frac{11\pi}{30}}\right]^n
\,\,\, ,
\EN
\begin{eqnarray}
& & U(x,x_1,\ldots,x_n) \,=\,\frac{1}{2}\,x^5\,\sum_{k_1,\ldots,k_6=0}^n
(-)^{k_1+k_3+k_5} \times
\label{pinc1}\\
& & \,\,\, x^{6n - (k_1+\cdots k_6)} \sin\left[\frac{\pi}{15}
\left(10 (k_1-k_2) + 6 (k_3-k_4) + (k_5-k_6)\right)\right] \,
\omega_{k_1}\ldots \omega_{k_6} \nonumber \, .
\eea

Note that for spinless operators, the total degree of the
polynomials $\Lambda_n$ is equal to $(3 n^2 - n)/2$. The partial degree of
$\Lambda_n$ in each variable $x_i$ is fixed by the asymptotic behaviour of the
operator which is considered. For the FF of the spin operator $\sigma(x)$,
the partial degree of $\Lambda_n$ is given by $(3 n -2)$.

Given the initial values of the recursive equations,
namely the one-particle FF $F_1$ and the two-particle FF $F_2$,
one can solve eqs. (\ref{pinc1}) and (\ref{pinc2}) in order to get
higher-particle FF. For instance, for the three-particle case we have
\EQ
\Lambda_3^{\Theta}(x_1,x_2,x_3) \,=\,-
{\cal A}_1 \,\omega_1 \omega_2 \,\left[ A
(\omega_1  \omega_2^4 + \omega_1^3 \omega_3^2) + B \omega_1^2
\omega_2^2 \omega_3 \right] \,\,\, ,
\EN
where
\[
A = \sin\frac{2\pi}{3} + \sin\frac{\pi}{15} + \sin\frac{2\pi}{5} \,\,\, ,
\]
\[
B =  3 \sin\frac{2\pi}{3} + 2\sin\frac{\pi}{15} +
2\sin\frac{2\pi}{5} + \sin\frac{2\pi}{15} -\sin\frac{\pi}{5} \,\,\, .
\]


\appsection

In this appendix we discuss some useful formulas for the calculation
of the functions $F^{min}(\th)$ defined in the text. Let us consider
a function $G_{\lambda}(\th)$, solution of the equations
\EQ
\begin{array}{cc}
G_{\lambda}(\theta ) & = f_{\lambda}(\th) G_{\lambda}(-\th) \,\,\, ;\\
G_{\lambda}(i \pi + \th) & = G_{\lambda}(i \pi - \th) \,\,\, ,
\end{array}
\EN
where
\EQ
f_{\lambda}(\th) = \frac{\tanh\frac{1}{2}\left(\th + i \pi \lambda\right)}
{\tanh\frac{1}{2}\left(\th - i \pi \lambda\right)}\,\,\, ,
\EN
without zeros and poles in the strip $(0, 2\pi i)$. This function admits
several different representations. One of them is given by
\EQ
G_{\lambda}(\th) = \exp\left[ 2 \int_0^{\infty} \frac{dt}{t}
\frac{\cosh\left[\left(\lambda - \frac{1}{2}\right) t\right]}
{\cosh\frac{t}{2} \sinh t}
\sin^2\frac{\hat\th t}{2 \pi}\right] \,\,\, ,
\label{integral}
\EN
where $\hat\th \equiv i\pi -\th$. This function presents an infinite number
of poles and zeros outside the strip $(0,2 \pi i)$, as shown by the following
infinite-product representation
\EQ
G_{\lambda}(\th) = \prod_{k=0}^{\infty}
\left[ \frac{
 \left[1 + \left(\frac{\hat\th/2\pi}{k+1 -\frac{\lambda}{2}}\right)^2
\right] \left[1 + \left(\frac{\hat\th/2\pi}{k+\frac{1}{2} + \frac{\lambda}{2}}
\right)^2\right]}
{\left[1 + \left(\frac{\hat\th/2\pi}{k +1 + \frac{\lambda}{2}}
\right)^2\right] \left[1 + \left(\frac{\hat\th/2\pi}{k+1 -
\frac{3}{2} - \frac{\lambda}{2}}\right)^2
\right]}\right]^{k+1} \,\,\, ,
\label{power}
\EN
or, equivalently
\EQ
G_\lambda(\th) = \prod_{k=0}^{\infty}
\left|\frac{\Gamma\left(\frac{1}{2} + k + \frac{\lambda}{2}\right)
\Gamma\left(1+k-\frac{\lambda}{2}\right) \Gamma\left(1+k+\frac{\lambda}{2} +
i \frac{\hat\th}{2\pi}\right) \Gamma\left(\frac{3}{2} + k -\frac{\lambda}{2} +
i \frac{\hat\th}{2\pi}\right)}
{\Gamma\left(\frac{3}{2} + k + \frac{\lambda}{2}\right)
\Gamma\left(1+k+\frac{\lambda}{2}\right) \Gamma\left(1+k-\frac{\lambda}{2} +
i \frac{\hat\th}{2\pi}\right) \Gamma\left(\frac{1}{2} + k +\frac{\lambda}{2} +
i \frac{\hat\th}{2\pi}\right)}\right|^2 \,\,\, .
\label{gamma}
\EN
For numerical calculation, a quite useful expression is given by the
mixed representation
\begin{eqnarray}
& & G_{\lambda}(\th) =
\prod_{k=0}^{N-1}
\left[ \frac{
\left[1 + \left(\frac{\hat\th/2\pi}{k+1 -\frac{\lambda}{2}}\right)^2
\right] \left[1 + \left(\frac{\hat\th/2\pi}{k+\frac{1}{2} + \frac{\lambda}{2}}
\right)^2\right] }
{\left[1 + \left(\frac{\hat\th/2\pi}{k +1 + \frac{\lambda}{2}}
\right)^2\right] \left[1 + \left(\frac{\hat\th/2\pi}{k+1 -
\frac{3}{2} - \frac{\lambda}{2}}\right)^2
\right]}\right]^{k+1} \times
\label{mixed}\\
& & \,\,\,\, \exp\left[ 2 \int_0^{\infty} \frac{dt}{t}
\frac{\cosh\left[\frac{t}{2} (1 - 2\lambda)\right]}{\cosh\frac{t}{2} \sinh t}
(N + 1 - N e^{-2t}) e^{-2 N t} \sin^2\frac{\hat\th t}{2 \pi}\right]\,\,\,.
\nonumber
\end{eqnarray}
{}From the above equations, we have
\EQ
G_{\lambda}(\th) = G_{1-\lambda}(\th) \,\,\, ,
\EN
and
\EQ
G_0(\th) = G_1(\th) = - i \sinh\frac{\th}{2} \,\,\, .
\EN
The function $G_{\lambda}(\th)$ satisfies the following functional equations
\EQ
G_{\lambda}(\th + i \pi) G_{\lambda}(\th) =
-i \frac{G_{\lambda}(0)}{\sin\pi \lambda} (\sinh\th + i \sin\pi \lambda)
\,\,\, ,
\label{pinchkin}
\EN
\EQ
G_{\lambda}(\th + i\pi \gamma) G_{\lambda}(\th - i\pi \gamma) =
\left(\frac{G_{\lambda}(i \pi \gamma) G_{\lambda}(-i\pi \gamma)}
{G_{\lambda+\gamma}(0) G_{\lambda - \gamma}(0)}\right)
G_{\lambda + \gamma}(\th) G_{\lambda - \gamma}(\th) \,\,\, ,
\label{pinchbound}
\EN
\EQ
G_{1-\lambda}(\th) G_{\lambda-1}(\th) =
\frac{\sinh\frac{1}{2}[\th - i (\lambda-1) \pi] \sinh\frac{1}{2}[\th + i
(\lambda+1) \pi]}
{\sin^2\frac{\pi \lambda}{2}} \,\,\, .
\label{cancellation}
\EN

\appsection

Aim of this Appendix is to illustrate through a specific example the
procedure used to obtain the results listed in Tables 3 and
4. Consider for instance the amplitude $S_{22}(\th)$. It exhibits
five simple positive residue poles located at $\th=i\frac{4\pi}{5}$,
$\th=i\frac{2\pi}{3}$, $\th=i\frac{7\pi}{15}$, $\th=i\frac{4\pi}{15}$ and
$\th=i\frac{\pi}{15}$ which correspond to the bound states $A_1$,
$A_2$, $A_4$, $A_5$ and $A_6$ respectively. In addition there is a
pair of crossing-symmetric double poles at $\th=i\frac{3\pi}{5}$ and
$\th=i\frac{2\pi}{5}$. The first one can be associated to the diagram
of Figure 9.a with $a=b=2$ and $c=d=e=1$ while the second one can be
associated to the diagram of Figure 9.b with $a=b=2$, $c=d=f=1$ and
$e=2$. A simple way to check that these diagrams actually satisfy the
required constraints (all the particles on mass-shell and
energy-momentum conservation) is to draw the vector diagram for the
momenta (dual diagram). Indeed, after performing the formal
substitution $\th_k\goto i\beta_k$, any two-momentum
$p_k^\mu=(m_k\cos\beta_k,im_k\sin\beta_k)$ can be thought as a vector
of length $m_k$ in the Euclidean plane. Then, the complete elasticity
of the process requires that the external momenta form a closed
parallelogram. Energy-momentum conservation also implies that the
momenta entering each three-particle vertex form closed triangles; as
a simple consequence of eq.(\ref{triangle}), the angle opposite to the
side of length $m_c$ must be $\overline u_{ab}^c\equiv
i\pi-u_{ab}^c$. If the triangles perfectly fit inside the external
parallelogram, the original space-time diagram satisfy all the
requirements. The dual diagrams associated to the double poles at
$\th=i\frac{3\pi}{5}$ and $\th=i\frac{2\pi}{5}$ in the amplitude
$S_{22}(\th)$ are shown in Figures 15.a and 15.b respectively. Of course,
the internal parallelogram in the first diagram corresponds to the
central interaction in Figure 9.a.

It can be easily checked that
\EQ
S_{22}\left(\th\simeq\frac{3 i
\pi}{5}\right)
\simeq\frac{(\Gamma_{11}^2)^4S_{11}
(\frac{i\pi}{5})}{(\th-\frac{3i\pi}{5})^2}\,\,\,,
\EN
\EQ
(\Gamma_{11}^2)^2 S_{11}\left(\frac{i\pi}{5}\right) = (\Gamma_{22}^1)^2\,\,\,.
\EN
According to the general discussion of Section 4, we write the FF
$F_{22}^\Theta(\th)$ as
\EQ
F_{22}^\Theta(\th)=Q_{22}^\Theta(\theta)\frac{F_{22}^{min}(\th)}{D_{22}(\th)}\,\,,
\EN
where
\EQ
F_{22}^{min}(\th) = -i\sinh\frac{\th}{2}\left[G_{\frac{4}{5}}G_{\frac{2}{3}}
G_{\frac{7}{15}}G_{\frac{4}{15}}G_{\frac{1}{15}}G_{\frac{3}{5}}^2\right](\th)
\,\,\,,
\EN
\EQ
D_{22}(\th) = \left[P_{\frac{4}{5}}P_{\frac{2}{3}}
P_{\frac{7}{15}}P_{\frac{4}{15}}P_{\frac{1}{15}}P_{\frac{3}{5}}
P_{\frac{2}{5}}\right](\th)
\,\,\,,
\EN
\EQ
Q_{22}^\Theta(\th)=\sum_{k=0}^{N_{22}}a_{22}^k\cosh^k\th\,\,\,.
\EN
Notice that
\[
F_{22}^{min}(\th) \sim e^{4|\th|}\,\,,\hspace{.8cm}|\th|\goto\infty
\,\,\, ,
\]
\[
D_{22}(\th) \sim e^{7|\th|}\,\,,\hspace{.8cm}|\th|\goto\infty \,\,\,.
\]
In view of these asymptotic behaviours, eqs.\,(\ref{bound}), (\ref{bbb})
imply
that $N_{22}\leq 3$. Since $F_{11}^{\Theta}$, $F_{12}^{\Theta}(\th)$ and
$F_a^\Theta$, $a=1,\ldots,5$ have been already determined in previous
steps of the bootstrap procedure, eqs.\,(\ref{ipi}), (\ref{pole}) and
(\ref{ffres2}) provide the following seven linear equations for the four
unknowns $a^k_{22}$, $k=0,\dots,3$
\EQ
F_{22}^\Theta(i\pi)=2\pi m_2^2\,\,\,,
\EN
\EQ
F_{22}^\Theta(\th\simeq
iu_{22}^c)\simeq i\frac{\Gamma_{22}^c}{\th-iu_{22}^c}\,F_{c}^\Theta\,\,,
\hspace{1cm}c=1,2,4,5\,\,\,,
\EN
\EQ
F_{22}^\Theta\left(\th\simeq
\frac{3i\pi}{5}\right) \simeq i\frac{(\Gamma_{11}^2)^2}
{\th-\frac{3i\pi}{5}}\,F_{11}^\Theta
\left(\frac{i\pi}{5}\right)\,\,\,,
\EN
\EQ
F_{22}^\Theta\left(\th\simeq
\frac{2i\pi}{5}\right) \simeq i\frac{\Gamma_{11}^2
\Gamma_{12}^2}{\th-\frac{2i\pi}{5}}\,
F_{12}^\Theta(0)\,\,\,.
\EN
These equations turn out to be compatible and the solution is contained
in Table 4. The particle $A_6$ appears for the first time as a bound
state in the amplitude $S_{22}(\th)$. Since $F_{22}^\Theta(\th)$ has
been fixed, $F_6^\Theta$ can now be extracted using again eq.\,(\ref{pole}).


\newpage

\hs

\vspace{25mm}

{\bf Table Caption}

\vspace{1cm}

\begin{description}
\item [Table 1]. $S$-matrix elements of the IMMF where the numbers $(\gamma)$
in parenthesis should be read in units of $\frac{1}{30}$. They
represent the functions $\tanh\frac{1}{2}
\left(\th+i\pi \frac{\gamma}{30}\right)/\tanh\frac{1}{2}
\left(\th-i\pi \frac{\gamma}{30}\right)$. The bound state poles related to the
particles $A_i$ ($i=1,\ldots,8$) are identified by the numbers $i$ placed
above the functions $(\gamma)$.
\item [Table 2]. The first seventeen lowest energy states entering the spectral
series.
\item [Table 3]. One-particle Form Factors of the operator $\Theta$.
\item [Table 4]. Coefficients of the polynomials $P_{ab}(\th)$.
\item [Table 5]. The first eight contributions to the $c$-theorem.
\item [Table 6]. The first eight contributions to the universal
amplitude of the free-energy.
\item [Table 7]. Numerical values of $G_c(x)$ and their errors $\delta$ for
$L=64$ and $h=0.075$, as determined in \cite{LR}.

\end{description}

\newpage

\hs

{\bf Figure Caption}

\vspace{5mm}

\begin{description}
\item [Figure 1]. Two-body $S$-matrix.
\item [Figure 2]. Analytic structure of the $S$-matrix in the $s$ plane.
The circles indicate the location of the bound state poles.
\item [Figure 3]. Bound state diagram in the $s$-channel.
\item [Figure 4]. Spectral representation series of the correlator.
\item [Figure 5]. Form Factor of the operator $\Phi$.
\item [Figure 6]. The two kinematical situations (a) and (b) entering
the residue equations of the annihilation poles.
\item [Figure 7]. Bound state bootstrap equation for the Form Factor.
\item [Figure 8]. Residue equation for the bound state poles of the
two-particle Form Factor.
\item [Figure 9]. Multi-scattering processes resulting in the double poles
of the $S$-matrix.
\item [Figure 10]. Residue equation for the simple poles of the Form Factor
associated to the double poles of the $S$-matrix.
\item [Figure 11]. Third-order pole in the $S$-matrix.
\item [Figure 12]. Residue equation for the double poles of the Form Factor
associated to the third-order poles of the $S$-matrix.
\item [Figure 13]. Correlation function $<\sigma(x) \sigma(0)>_c$ versus
lattice space distances. The points on the graph represent the numerical
data, as extracted from Table 7, while the continuum curve is the theoretical
estimate  with only the first three Form Factors.
\item [Figure 14]. Correlation function $<\sigma(x) \sigma(0)>_c$ versus
lattice space distances. The points on the graph represent the numerical
data, as extracted from Table 7 while the continuum curve is the theoretical
estimate obtained with the first eight terms of the spectral series.
\item [Figure 15]. Dual diagrams of double pole graphs.
\end{description}


\begin{center}

\vspace{3cm}
\begin{tabular}{|c|c|}\hline
$a$ \,\, $b$ &
$S_{ab}$ \\ \hline \hline
1 \,\, 1 &
$ \st{\bf 1}{(20)} \, \st{\bf 2}{(12)} \, \st{\bf 3}{(2)} $\\ \hline
1 \,\, 2 &
$ \st{\bf 1}{(24)} \, \st{\bf 2}{(18)} \, \st{\bf 3}{(14)} \, \st{\bf 4}{(8)}
$\\ \hline
1 \,\, 3 &
$ \st{\bf 1}{(29)} \, \st{\bf 2}{(21)} \, \st{\bf 4}{(13)} \,
\st{\bf 5}{(3)} \, (11)^2 $ \\ \hline
1 \,\, 4 &
$ \st{\bf 2}{(23)} \, \st{\bf 3}{(21)} \, \st{\bf 4}{(17)} \,
\st{\bf 5}{(11)} \, \st{\bf 6}{(7)} \, (15) $ \\ \hline
1 \,\, 5 &
$ \st{\bf 3}{(28)} \, \st{\bf 4}{(22)} \, \st{\bf 6}{(14)} \,
\st{\bf 7}{(4)} \, (10)^2 \, (12)^2 $ \\ \hline
1 \,\, 6 &
$ \st{\bf 4}{(25)} \, \st{\bf 5}{(19)} \, \st{\bf 7}{(9)} \,
(7)^2 \, (13)^2 \, (15) $ \\ \hline
1 \,\, 7 &
$ \st{\bf 5}{(27)} \, \st{\bf 6}{(23)} \, \st{\bf 8}{(5)} \,
(9)^2 \, (11)^2\, (13)^2 \, (15) $ \\ \hline
1 \,\, 8 &
$ \st{\bf 7}{(26)} \, \st{\bf 8}{(16)^3} \, (6)^2 \,
(8)^2 \, (10)^2 \, (12)^2 $ \\ \hline
2 \,\, 2 &
$ \st{\bf 1}{(24)} \, \st{\bf 2}{(20)} \, \st{\bf 4}{(14)} \,
\st{\bf 5}{(8)} \,\st{\bf 6}{(2)} \, (12)^2 $ \\ \hline
2 \,\, 3 &
$ \st{\bf 1}{(25)} \, \st{\bf 3}{(19)} \, \st{\bf 6}{(9)} \, (7)^2 \,
(13)^2 \, (15) $ \\ \hline
2 \,\, 4 &
$ \st{\bf 1}{(27)} \,\st{\bf 2}{(23)} \, \st{\bf 7}{(5)} \,
(9)^2 \, (11)^2 \, (13)^2\,(15)$ \\ \hline
2 \,\, 5 &
$ \st{\bf 2}{(26)} \,\st{\bf 6}{(16)^3} \, (6)^2 (8)^2 (10)^2 (12)^2 $
\\ \hline
2 \,\, 6 &
$ \st{\bf 2}{(29)} \, \st{\bf 3}{(25)} \, \st{\bf 5}{(19)^3} \,
\st{\bf 7}{(13)^3} \, \st{\bf 8}{(3)} \, (7)^2 (9)^2 (15) $ \\ \hline
2 \,\,7 &
$ \st{\bf 4}{(27)} \, \st{\bf 6}{(21)^3} \, \st{\bf 7}{(17)^3} \,
\st{\bf 8}{(11)^3} \, (5)^2 (7)^2 (15)^2 $ \\ \hline
2 \,\, 8 &
$ \st{\bf 6}{(28)} \, \st{\bf 7}{(22)^3} \, (4)^2 (6)^2 (10)^4 (12)^4 (16)^4 $
\\ \hline
3 \,\, 3 &
$ \st{\bf 2}{(22)} \, \st{\bf 3}{(20)^3} \, \st{\bf 5}{(14)} \,
\st{\bf 6}{(12)^3} \, \st{\bf 7}{(4)} \, (2)^2 $ \\ \hline
3 \,\, 4 &
$ \st{\bf 1}{(26)} \, \st{\bf 5}{(16)^3} \, (6)^2 (8)^2 (10)^2 (12)^2 $
\\ \hline
3 \,\, 5 &
$ \st{\bf 1}{(29)} \, \st{\bf 3}{(23)} \, \st{\bf 4}{(21)^3} \,
\st{\bf 7}{(13)^3} \, \st{\bf 8}{(5)} \, (3)^2 (11)^4 (15) $ \\ \hline
3 \,\, 6 &
$ \st{\bf 2}{(26)} \, \st{\bf 3}{(24)^3} \, \st{\bf 6}{(18)^3} \,
\st{\bf 8}{(8)^3} \, (10)^2 (16)^4 $ \\ \hline
3 \,\, 7 &
$ \st{\bf 3}{(28)} \, \st{\bf 5}{(22)^3} \, (4)^2 (6)^2 (10)^4 (12)^4 (16)^4
$ \\ \hline
3 \,\, 8 &
$ \st{\bf 5}{(27)} \, \st{\bf 6}{(25)^3} \, \st{\bf 8}{(17)^5} \,
(7)^4 (9)^4 (11)^2 (15)^3 $ \\ \hline
\end{tabular}
\end{center}

\begin{center}
Continued
\end{center}

\newpage
\begin{center}

\vspace{3cm}
\begin{tabular}{|c|c|}\hline
$a$ \,\, $b$ &
$S_{ab}$ \\ \hline \hline
4 \,\, 4 &
$ \st{\bf 1}{(26)} \, \st{\bf 4}{(20)^3} \, \st{\bf 6}{(16)^3} \,
\st{\bf 7}{(12)^3} \, \st{\bf 8}{(2)} \, (6)^2 (8)^2 $ \\ \hline
4 \,\, 5 &
$ \st{\bf 1}{(27)} \, \st{\bf 3}{(23)^3} \, \st{\bf 5}{(19)^3} \,
\st{\bf 8}{(9)^3} \, (5)^2 (13)^4 (15)^2 $ \\ \hline
4 \,\, 6 &
$ \st{\bf 1}{(28)} \, \st{\bf 4}{(22)^3} (4)^2 (6)^2 (10)^4
(12)^4 (16)^4 $ \\ \hline
4 \,\, 7 &
$ \st{\bf 2}{(28)} \, \st{\bf 4}{(24)^3} \, \st{\bf 7}{(18)^5} \,
\st{\bf 8}{(14)^5} \, (4)^2 (8)^4 (10)^4 $ \\ \hline
4 \,\, 8 &
$ \st{\bf 4}{(29)} \, \st{\bf 5}{(25)^3} \, \st{\bf 7}{(21)^5} \,
(3)^2 (7)^4 (11)^6 (13)^6 (15)^3 $ \\ \hline
5 \,\, 5 &
$ \st{\bf 4}{(22)^3} \, \st{\bf 5}{(20)^5} \, \st{\bf 8}{(12)^5} \,
(2)^2 (4)^2 (6)^2 (16)^4 $ \\ \hline
5 \,\, 6 &
$ \st{\bf 1}{(27)} \, \st{\bf 2}{(25)^3} \, \st{\bf 7}{(17)^5} \,
(7)^4 (9)^4 (11)^4 (15)^3 $ \\ \hline
5 \,\, 7 &
$ \st{\bf 1}{(29)} \, \st{\bf 3}{(25)^3} \, \st{\bf 6}{(21)^5} \,
(3)^2 (7)^4 (11)^6 (13)^6 (15)^3 $ \\ \hline
5 \,\, 8 &
$ \st{\bf 3}{(28)} \, \st{\bf 4}{(26)^3} \, \st{\bf 5}{(24)^5} \,
\st{\bf 8}{(18)^7} \, (8)^6 (10)^6 (16)^8 $ \\ \hline
6 \,\, 6 &
$ \st{\bf 3}{(24)^3} \, \st{\bf 6}{(20)^5} \, \st{\bf 8}{(14)^5} \,
(2)^2 (4)^2 (8)^4 (12)^6 $ \\ \hline
6 \,\, 7 &
$ \st{\bf 1}{(28)} \, \st{\bf 2}{(26)^3} \, \st{\bf 5}{(22)^5} \,
\st{\bf 8}{(16)^7} \, (6)^4 (10)^6 (12)^6 $ \\ \hline
6 \,\, 8 &
$ \st{\bf 2}{(29)} \, \st{\bf 3}{(27)^3} \, \st{\bf 6}{(23)^5} \,
\st{\bf 7}{(21)^7} \, (5)^4 (11)^8 (13)^8 (15)^4 $ \\ \hline
7 \,\, 7 &
$ \st{\bf 2}{(26)^3} \, \st{\bf 4}{(24)^5} \, \st{\bf 7}{(20)^7} \,
(2)^2 (8)^6 (12)^8 (16)^8 $ \\ \hline
7 \,\, 8 &
$ \st{\bf 1}{(29)} \, \st{\bf 2}{(27)^3} \, \st{\bf 4}{(25)^5} \,
\st{\bf 6}{(23)^7} \, \st{\bf 8}{(19)^9} \, (9)^8 (13)^{10} (15)^ 5 $
\\ \hline
8 \,\, 8 &
$ \st{\bf 1}{(28)^3} \, \st{\bf 3}{(26)^5} \, \st{\bf 5}{(24)^7} \,
\st{\bf 7}{(22)^9} \, \st{\bf 8}{(20)^{11}} \, (12)^{12} (16)^{12} $ \\
\hline
\end{tabular}
\end{center}

\begin{center}
{\bf Table 1}
\end{center}

\newpage

\begin{center}

\vspace{3cm}
\begin{tabular}{|c|c|}\hline
$|n> $& $\sqrt{s} $\\
\hline \hline
$ A_1$  & $ m_1$\\ \hline
$ A_2 $  & $(1.6180..) \, m_1 $\\ \hline
$ A_3 $  &  $(1.9890..) \, m_1 $ \\ \hline
$ A_1 A_1 $ & $ \geq 2 \, m_1 $  \\ \hline
$ A_4 $  & $ (2.4048..) m_1$ \\ \hline
$ A_1 A_2 $  & $ \geq (2.6180..) m_1$ \\ \hline
$ A_5 $  & $ (2.9563..) m_1$ \\ \hline
$ A_1 A_3 $  & $ \geq (2.9890..) m_1$  \\ \hline
$ A_1 A_1 A_1 $ & $ \geq 3 m_1$ \\ \hline
$ A_6 $  & $ (3.2183..) m_1$ \\ \hline
$ A_2 A_2 $  & $ \geq (3.2360..) m_1$  \\ \hline
$ A_1 A_4 $  & $ \geq (3.4048..) m_1$  \\ \hline
$ A_2 A_3 $  & $ \geq (3.6070..) m_1$  \\ \hline
$ A_1 A_1 A_2 $ & $\geq (3.6180..) m_1$ \\ \hline
$ A_7 $ & $ 3.8911.. m_1$ \\ \hline
$ A_1 A_5 $ & $ \geq (3.9563..) m_1$ \\ \hline
$ A_3 A_3 $ & $ \geq (3.9780..) m_1$  \\ \hline
\end{tabular}
\end{center}

\begin{center}
{\bf Table 2}
\end{center}

\newpage

\begin{center}

\vspace{3cm}
\begin{tabular}{|c|r|}\hline
$ F^{\Theta}_1 $ &
0.4971505471  \\
$ F^{\Theta}_2$ &
- 0.2627111760  \\
$ F^{\Theta}_3 $ &
0.1447685755  \\
$ F^{\Theta}_4 $ &
- 0.1107486745  \\
$ F^{\Theta}_5 $ &
- 0.0467951944  \\
$ F^{\Theta}_6 $ &
0.0336573286  \\
$ F^{\Theta}_7 $ &
- 0.0127414814  \\
$ F^{\Theta}_8 $ &
0.0023550931  \\ \hline
\end{tabular}
\end{center}

\begin{center}
{\bf Table 3}
\end{center}

\newpage

\begin{center}

\vspace{3cm}
\begin{tabular}{|c|}\hline

$ a_{11}^1 = 1.623628945  $ \\
$ a_{11}^0 = 7.906814252  $ \\ \hline
$ a_{12}^1 = 6.189362113  $ \\
$ a_{12}^0 = 48.76949783  $\\ \hline
$ a_{13}^2 = 451.6582994  $\\
$ a_{13}^1 = 4824.093743  $ \\
$ a_{13}^0 = 4389.138083  $ \\ \hline
$ a_{22}^3 = 16.66647213  $ \\
$ a_{22}^2 = 258.9398602  $ \\
$ a_{22}^1 = 613.8726746  $ \\
$ a_{22}^0 = 388.0488792  $ \\ \hline
$ a_{14}^2 =-17.51368456  $ \\
$ a_{14}^1 =-222.5261059  $ \\
$ a_{14}^0 =-207.6061104  $ \\ \hline
$ a_{23}^3 = 71.93457181  $ \\
$ a_{23}^2 = 1358.926475  $ \\
$ a_{23}^1 = 3745.854658  $ \\
$ a_{23}^0 = 2718.541527  $ \\ \hline
$ a_{15}^3 = 202.2760111  $ \\
$ a_{15}^2 = 3328.024309  $ \\
$ a_{15}^1 = 6286.863589  $ \\
$ a_{15}^0 = 3162.939632  $ \\ \hline
$ a_{33}^5 = 928.5620526  $ \\
$ a_{33}^4 = 23400.66614  $ \\
$ a_{33}^3 = 116753.1311  $ \\
$ a_{33}^2 = 233559.8778  $ \\
$ a_{33}^1 = 207377.6722  $ \\
$ a_{33}^0 = 68123.67968  $ \\ \hline
\end{tabular}

\end{center}

\begin{center}
{\bf Table 4}
\end{center}

\newpage

\, \, \,
\vspace{3cm}

\begin{center}

\vspace{3cm}
\begin{tabular}{|l|c|}\hline

$C_1 $ & 0.472038282  \\
$C_2 $ & 0.019231268 \\
$C_3 $ & 0.002557246 \\
$C_{11}$ & 0.003919717 \\
$C_4 $ & 0.000700348 \\
$C_{12}$ & 0.000974265 \\
$C_5 $ & 0.000054754 \\
$C_{13}$ & 0.000154186 \\ \hline
$C_{\rm partial}$ & 0.499630066 \\ \hline
\end{tabular}

\vspace{1cm}

\begin{center}
{\bf Table 5}
\end{center}

\newpage

\,\,\,

\vspace{6cm}
\begin{tabular}{|l|c|}\hline

$U_1 $ & 0.050084817 \\
$U_2 $ & 0.005342101 \\
$U_3 $ & 0.001073471 \\
$U_{11}$ & 0.002496286 \\
$U_4 $ & 0.000429759 \\
$U_{12}$ & 0.001031748 \\
$U_5 $ & 0.000050773 \\
$U_{13}$ & 0.000225505 \\ \hline
$U_{\rm partial}$ & 0.060734461 \\ \hline
\end{tabular}
\end{center}

\vspace{1cm}

\begin{center}
{\bf Table 6}
\end{center}

\newpage

\begin{center}

\begin{tabular}{|c|c|c|}\hline
$ x $ &
$ G(x) $ &
$ \delta $ \\
\hline \hline
$ 1 $ &
$ 0.210791 $&
$0.000045 $\\ \hline
$ 2 $ &
$ 0.118830 $ &
$ 0.000035 $ \\ \hline
$ 3 $ &
$ 0.076439 $ &
$ 0.000029 $ \\ \hline
$ 4 $ &
$ 0.052763 $ &
$ 0.000024 $ \\ \hline
$ 5 $ &
$ 0.037944 $ &
$ 0.000021 $ \\ \hline
$ 6 $ &
$ 0.028009 $ &
$ 0.000018 $ \\ \hline
$ 7 $ &
$ 0.021060 $ &
$ 0.000016 $ \\ \hline
$ 8 $ &
$ 0.016049 $ &
$ 0.000013 $ \\ \hline
$ 9 $ &
$ 0.012356 $ &
$ 0.000012 $ \\ \hline
$ 10   $ &
$ 0.009587 $ &
$ 0.000011 $ \\ \hline
$ 11 $ &
$ 0.007493 $ &
$ 0.000010 $ \\ \hline
$ 12 $ &
$ 0.005887 $ &
$ 0.000008 $ \\ \hline
$ 13 $ &
$ 0.004645 $ &
$ 0.000009 $ \\ \hline
$ 14 $ &
$ 0.003683 $ &
$ 0.000008 $ \\ \hline
$ 15 $ &
$ 0.002926 $ &
$ 0.000008 $ \\ \hline
$ 16 $ &
$ 0.002332 $ &
$ 0.000008 $ \\ \hline
$ 17 $ &
$ 0.001856 $ &
$ 0.000007 $ \\ \hline
$ 18 $ &
$ 0.001488 $ &
$ 0.000006 $ \\ \hline
$ 19 $ &
$ 0.001197 $ &
$ 0.000006 $ \\ \hline
$ 20 $ &
$ 0.000962 $ &
$ 0.000007 $ \\ \hline
$ 21 $ &
$ 0.000775 $ &
$ 0.000008 $ \\ \hline
$ 22 $ &
$ 0.000626 $ &
$ 0.000007 $ \\ \hline
$ 23 $ &
$ 0.000504 $ &
$ 0.000006 $ \\ \hline
$ 24 $ &
$ 0.000410 $ &
$ 0.000006 $ \\ \hline
$ 25 $ &
$ 0.000335 $ &
$ 0.000004 $ \\ \hline
$ 26 $ &
$ 0.000272 $ &
$ 0.000004 $ \\ \hline
$ 27 $ &
$ 0.000222 $ &
$ 0.000004 $ \\ \hline
$ 28 $ &
$ 0.000181 $ &
$ 0.000005 $ \\ \hline
$ 29 $ &
$ 0.000146 $ &
$ 0.000006 $ \\ \hline
$ 30 $ &
$ 0.000119 $ &
$ 0.000005 $ \\ \hline
$ 31 $ &
$ 0.000098 $ &
$ 0.000004 $ \\ \hline
$ 32 $ &
$ 0.000080 $ &
$ 0.000003 $ \\ \hline
\end{tabular}
\end{center}

\begin{center}
{\bf Table 7}
\end{center}

\samepage
\vspace*{-1cm}


\pagestyle{empty}

\begin{figure}
\hspace{-1cm}
\vspace{3cm}
\centerline{
\psfig{figure=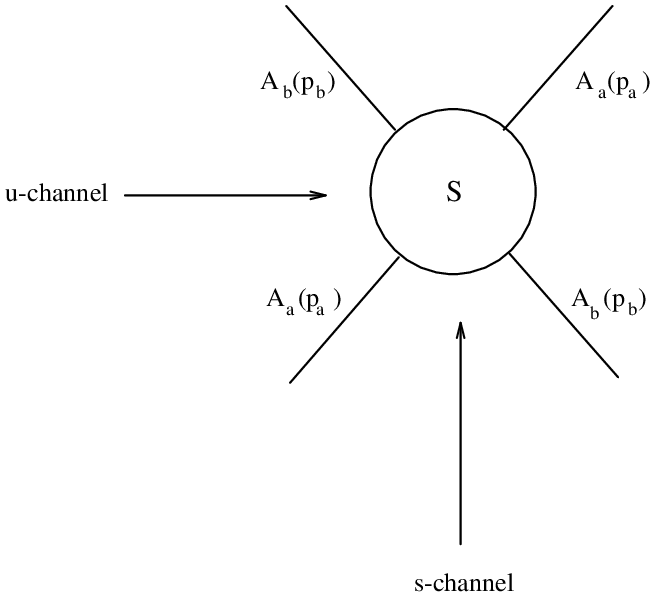}}
\vspace{3cm}
\begin{center}
\hspace{2cm}
{\bf Figure 1}
\end{center}
\end{figure}

\newpage
\begin{figure}
\centerline{
\psfig{figure=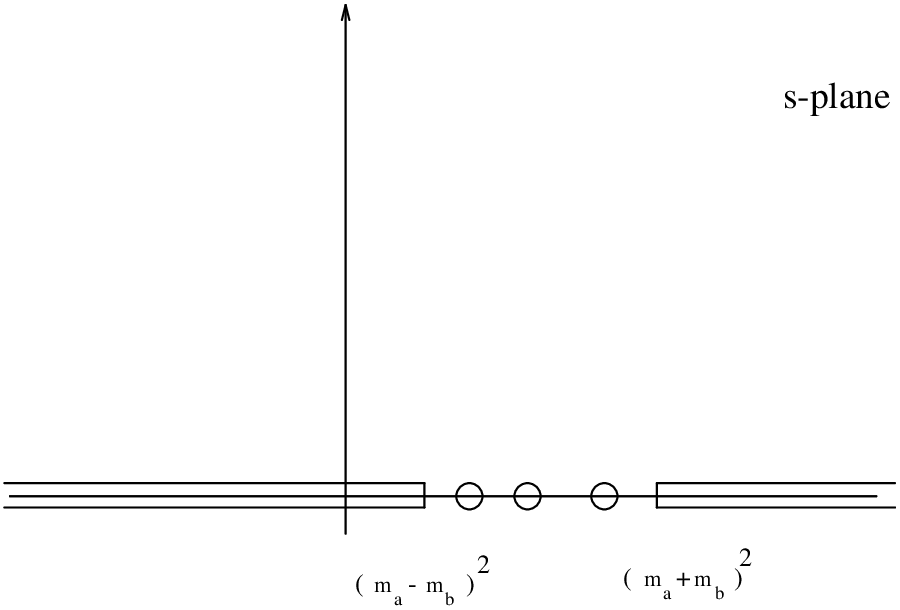}}
\vspace{3cm}
\begin{center}
{\bf Figure 2}
\end{center}
\end{figure}

\newpage

\newpage

\begin{figure}
\centerline{
\psfig{figure=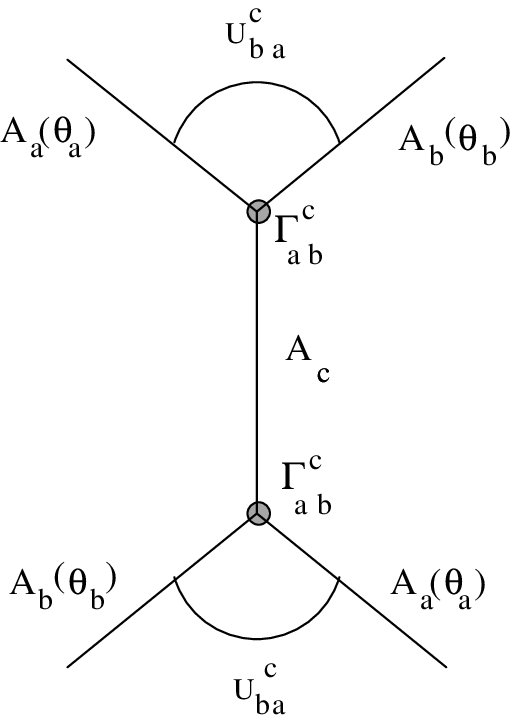}}
\vspace{3cm}
\begin{center}
{\bf Figure 3}
\end{center}
\end{figure}

\begin{figure}
\centerline{
\psfig{figure=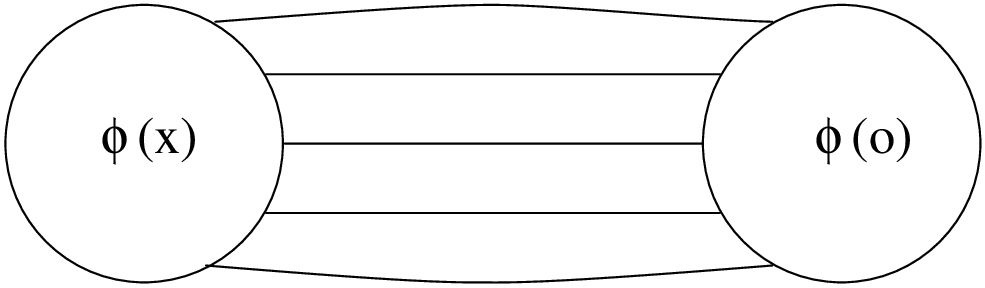}}
\vspace{3cm}
\begin{center}
{\bf Figure 4}
\end{center}
\end{figure}

\newpage

\begin{figure}
\centerline{
\psfig{figure=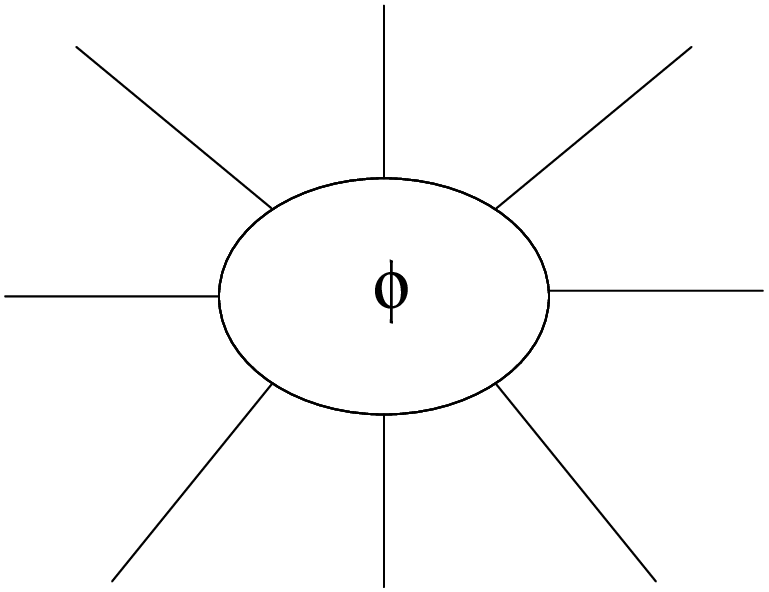}}
\vspace{3cm}
\begin{center}
{\bf Figure 5}
\end{center}
\end{figure}

\newpage

\begin{figure}
\centerline{
\psfig{figure=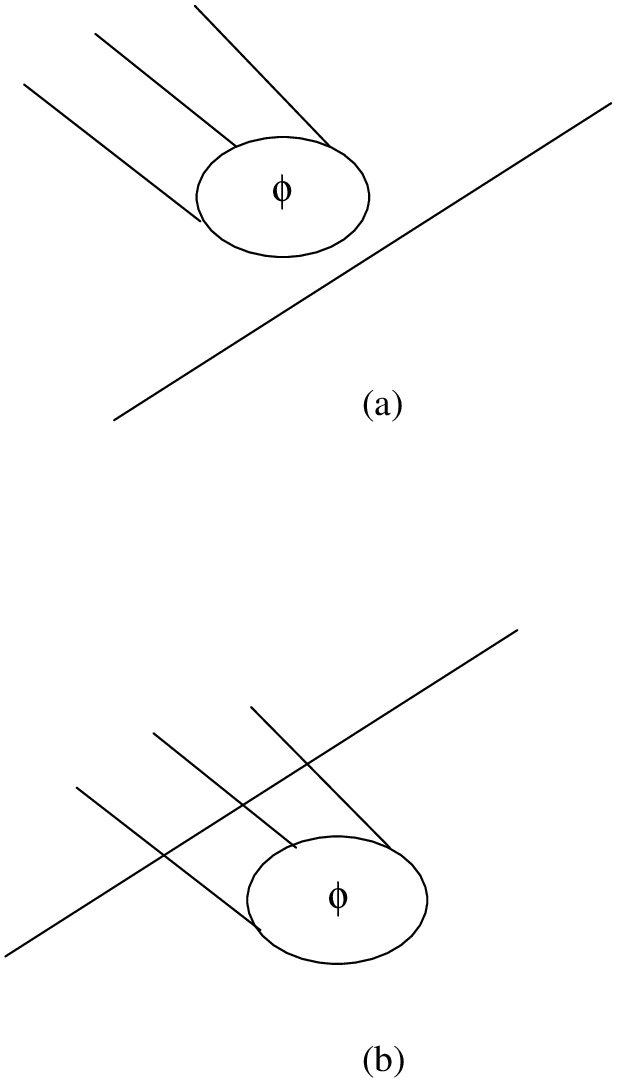}}
\vspace{3cm}
\begin{center}
{\bf Figure 6}
\end{center}
\end{figure}

\begin{figure}
\centerline{
\psfig{figure=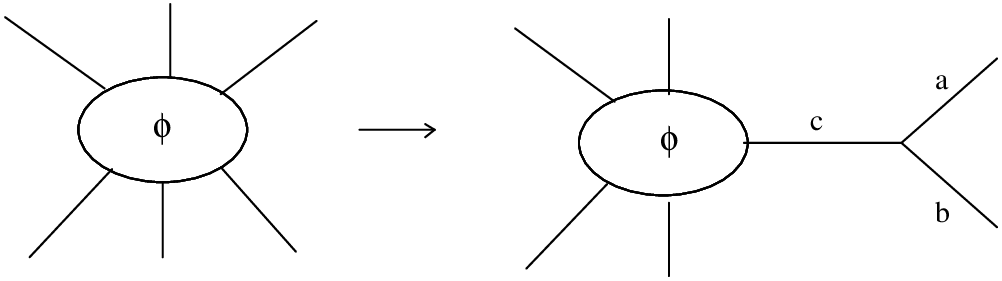}}
\vspace{3cm}
\begin{center}
{\bf Figure 7}
\end{center}
\end{figure}

\newpage

\begin{figure}
\centerline{
\psfig{figure=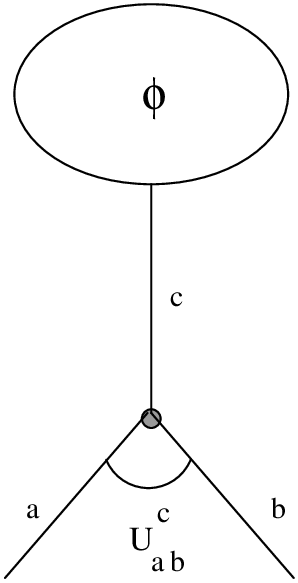}}
\vspace{3cm}
\begin{center}
{\bf Figure 8}
\end{center}
\end{figure}

\newpage

\begin{figure}
\centerline{
\psfig{figure=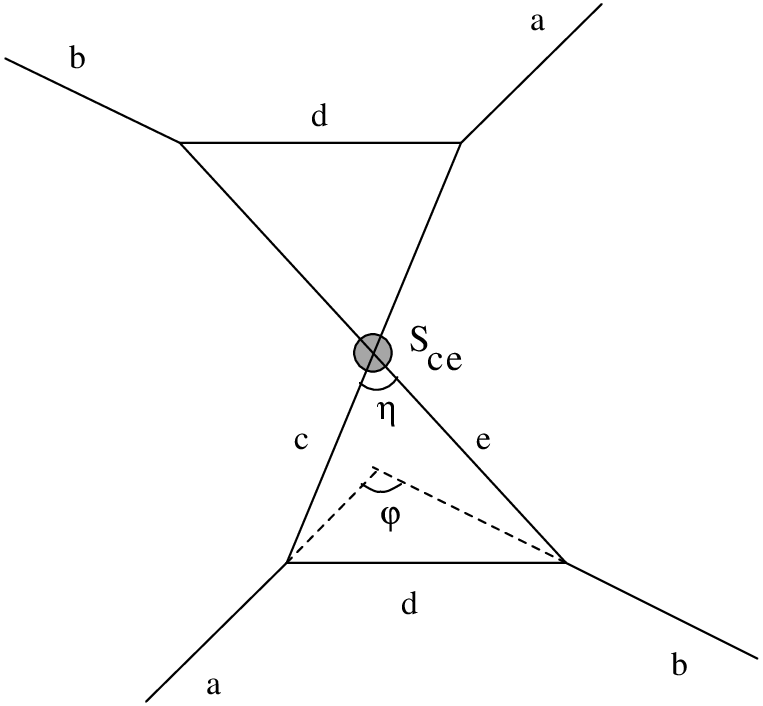}}
\vspace{3cm}
\begin{center}
{\bf Figure 9.a}
\end{center}
\end{figure}

\newpage

\begin{figure}
\centerline{
\psfig{figure=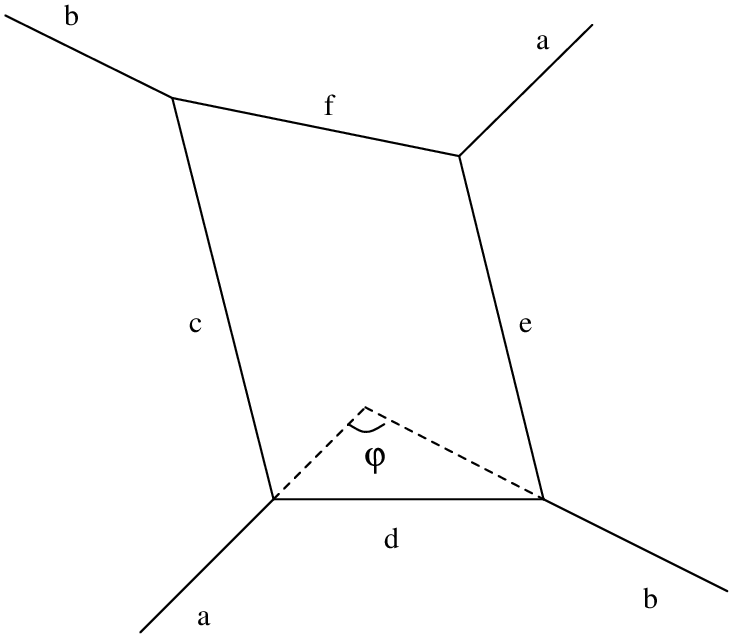}}
\vspace{3cm}
\begin{center}
{\bf Figure 9.b}
\end{center}
\end{figure}

\newpage

\begin{figure}
\centerline{
\psfig{figure=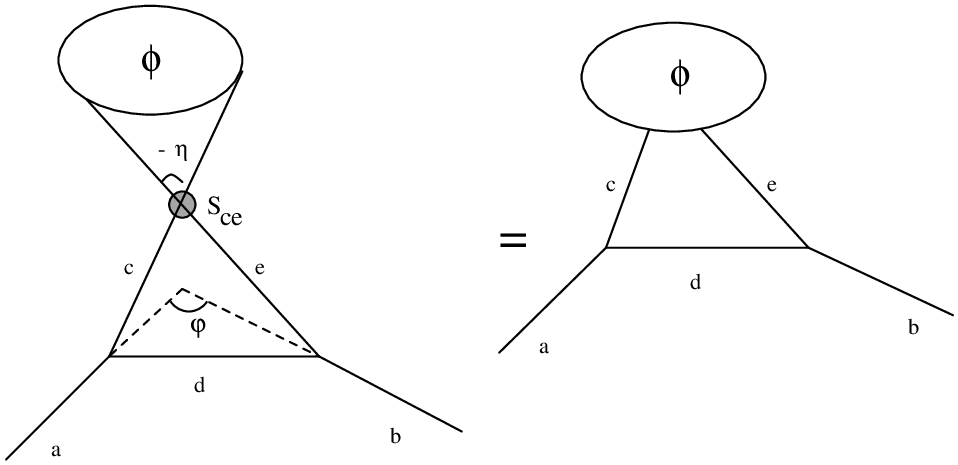}}
\vspace{3cm}

\newpage

\begin{center}
{\bf Figure 10}
\end{center}
\end{figure}

\newpage

\begin{figure}
\centerline{
\psfig{figure=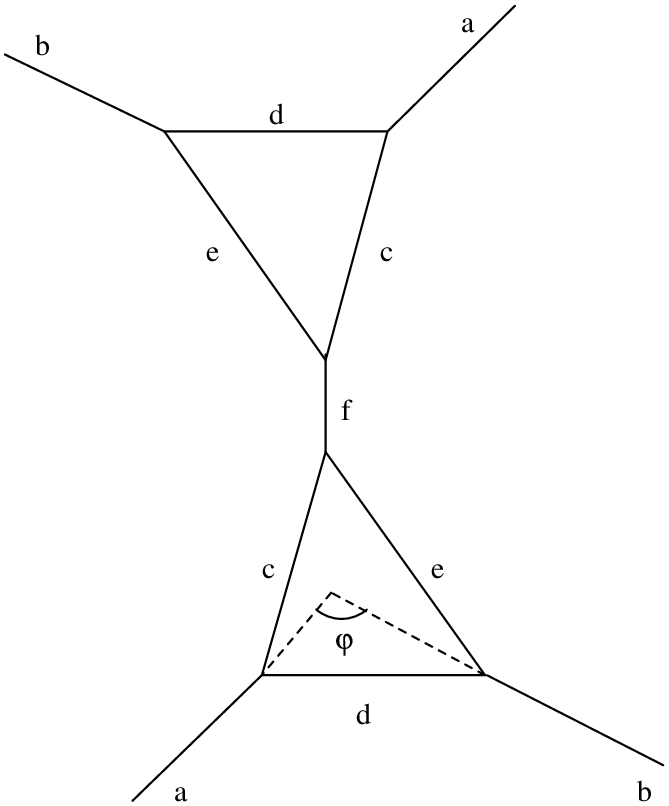}}
\vspace{3cm}
\begin{center}
{\bf Figure 11}
\end{center}
\end{figure}

\newpage

\begin{figure}
\centerline{
\psfig{figure=figure12.ps}}
\vspace{3cm}
\begin{center}
{\bf Figure 12}
\end{center}
\end{figure}

\newpage

\begin{figure}
\centerline{
\psfig{figure=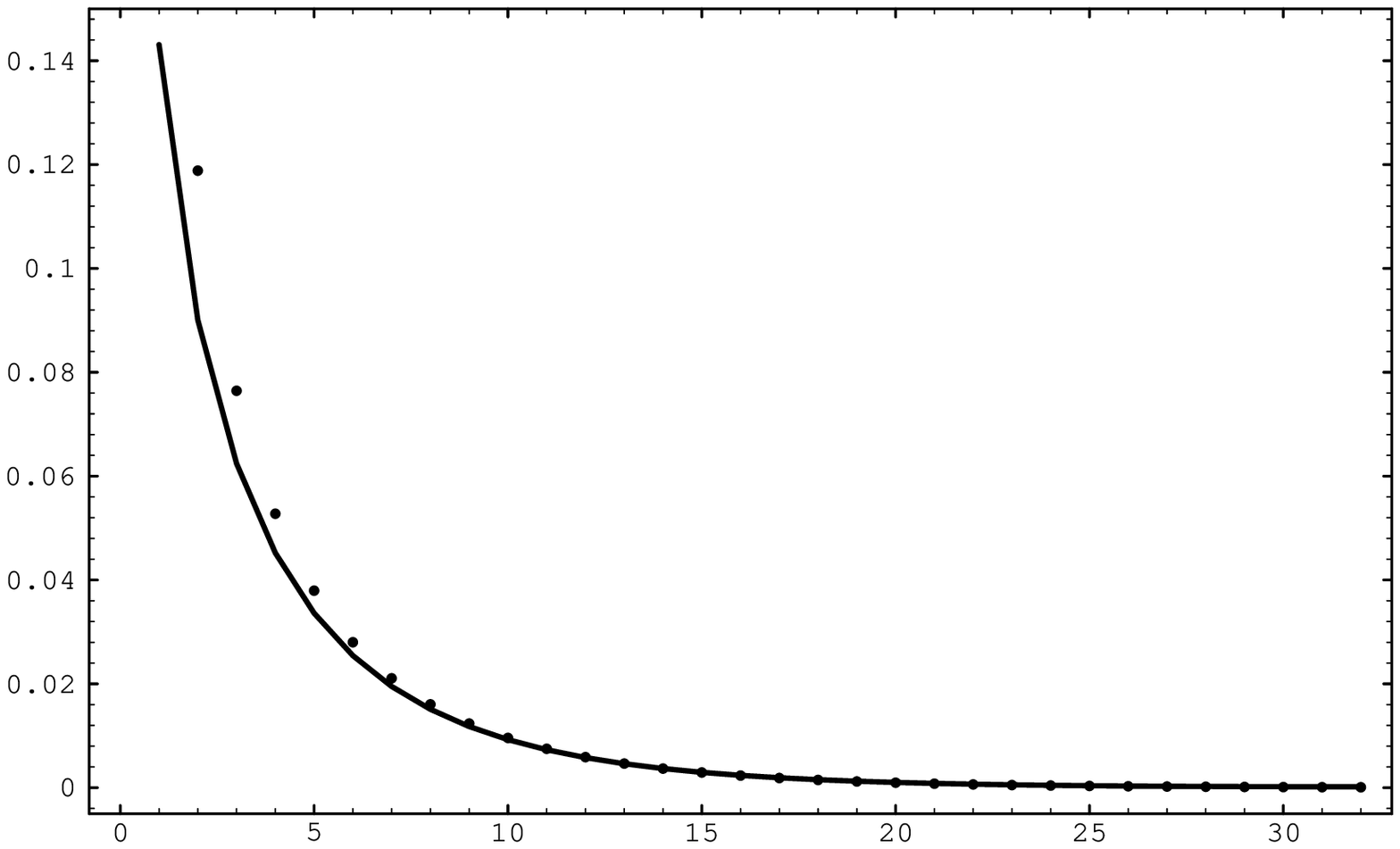}}
\vspace{-1cm}
\begin{center}
{\bf Figure 13}
\end{center}
\end{figure}

\newpage

\begin{figure}
\centerline{
\psfig{figure=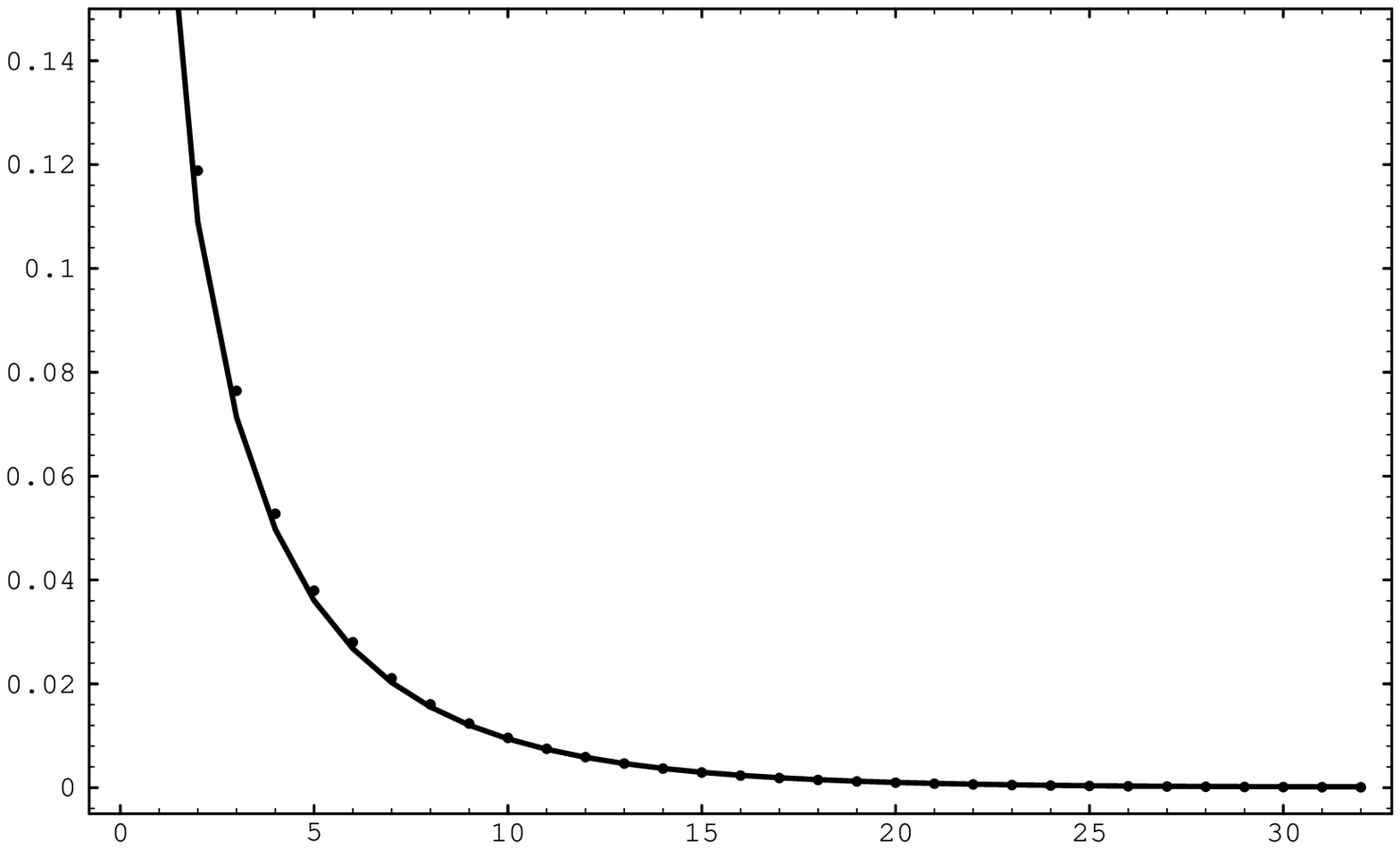}}
\vspace{-1cm}
\begin{center}
{\bf Figure 14}
\end{center}
\end{figure}

\newpage

\begin{figure}
\centerline{
\psfig{figure=figure15.ps}}
\vspace{3cm}
\begin{center}
{\bf Figure 15}
\end{center}
\end{figure}

\end{document}